\documentclass[twocolumn,superscriptaddress,prd,floatfix]{revtex4}
\usepackage{graphicx}
\usepackage{subfigure}
\usepackage{dcolumn}
\usepackage{bm}
\usepackage{mathtools,slashed}
\usepackage{amsmath}
\usepackage[usenames]{color}
\usepackage{array} 

\def\m{M}
\def\l{L}

\def\lsim{\mathrel{\rlap{\lower4pt\hbox{\hskip1pt$\sim$}}
   \raise1pt\hbox{$<$}}}
\def\gsim{\mathrel{\rlap{\lower4pt\hbox{\hskip1pt$\sim$}}
   \raise1pt\hbox{$>$}}}

\begin{document}
\title{Photoproduction of $K^+K^-$ meson pairs on the proton\\}

\newcommand*{\ANL}{Argonne National Laboratory, Argonne, Illinois 60439}
\newcommand*{\ANLindex}{1}
\affiliation{\ANL}
\newcommand*{\ASU}{Arizona State University, Tempe, Arizona 85287-1504}
\newcommand*{\ASUindex}{2}
\affiliation{\ASU}
\newcommand*{\CSUDH}{California State University, Dominguez Hills, Carson, CA 90747}
\newcommand*{\CSUDHindex}{3}
\affiliation{\CSUDH}
\newcommand*{\CANISIUS}{Canisius College, Buffalo, NY}
\newcommand*{\CANISIUSindex}{4}
\affiliation{\CANISIUS}
\newcommand*{\CMU}{Carnegie Mellon University, Pittsburgh, Pennsylvania 15213}
\newcommand*{\CMUindex}{5}
\affiliation{\CMU}
\newcommand*{\CUA}{Catholic University of America, Washington, D.C. 20064}
\newcommand*{\CUAindex}{6}
\affiliation{\CUA}
\newcommand*{\SACLAY}{IRFU, CEA, Universit'e Paris-Saclay, F-91191 Gif-sur-Yvette, France}
\newcommand*{\SACLAYindex}{7}
\affiliation{\SACLAY}
\newcommand*{\CNU}{Christopher Newport University, Newport News, Virginia 23606}
\newcommand*{\CNUindex}{8}
\affiliation{\CNU}
\newcommand*{\UCONN}{University of Connecticut, Storrs, Connecticut 06269}
\newcommand*{\UCONNindex}{9}
\affiliation{\UCONN}
\newcommand*{\DUKE}{Duke University, Durham, North Carolina 27708-0305}
\newcommand*{\DUKEindex}{10}
\affiliation{\DUKE}
\newcommand*{\FU}{Fairfield University, Fairfield CT 06824}
\newcommand*{\FUindex}{11}
\affiliation{\FU}
\newcommand*{\FERRARAU}{Universita' di Ferrara , 44121 Ferrara, Italy}
\newcommand*{\FERRARAUindex}{12}
\affiliation{\FERRARAU}
\newcommand*{\FIU}{Florida International University, Miami, Florida 33199}
\newcommand*{\FIUindex}{13}
\affiliation{\FIU}
\newcommand*{\FSU}{Florida State University, Tallahassee, Florida 32306}
\newcommand*{\FSUindex}{14}
\affiliation{\FSU}
\newcommand*{\Genova}{Universit$\grave{a}$ di Genova, 16146 Genova, Italy}
\newcommand*{\Genovaindex}{15}
\affiliation{\Genova}
\newcommand*{\GWUI}{The George Washington University, Washington, DC 20052}
\newcommand*{\GWUIindex}{16}
\affiliation{\GWUI}
\newcommand*{\ISU}{Idaho State University, Pocatello, Idaho 83209}
\newcommand*{\ISUindex}{17}
\affiliation{\ISU}
\newcommand*{\INDIANA} {Physics Department and Nuclear Theory Center \\ Indiana University, Bloomington, Indiana 47405}
\affiliation{\INDIANA}
\newcommand*{\IND} {Center for Exploration of Energy and Matter \\Indiana University, Bloomington, Indiana, 47403}
\affiliation{\IND}
\newcommand*{\INFNFE}{INFN, Sezione di Ferrara, 44100 Ferrara, Italy}
\newcommand*{\INFNFEindex}{18}
\affiliation{\INFNFE}
\newcommand*{\INFNFR}{INFN, Laboratori Nazionali di Frascati, 00044 Frascati, Italy}
\newcommand*{\INFNFRindex}{19}
\affiliation{\INFNFR}
\newcommand*{\INFNGE}{INFN, Sezione di Genova, 16146 Genova, Italy}
\newcommand*{\INFNGEindex}{20}
\affiliation{\INFNGE}
\newcommand*{\INFNRO}{INFN, Sezione di Roma Tor Vergata, 00133 Rome, Italy}
\newcommand*{\INFNROindex}{21}
\affiliation{\INFNRO}
\newcommand*{\INFNTUR}{INFN, Sezione di Torino, 10125 Torino, Italy}
\newcommand*{\INFNTURindex}{22}
\affiliation{\INFNTUR}
\newcommand*{\ORSAY}{Institut de Physique Nucl\'eaire, CNRS/IN2P3 and Universit\'e Paris Sud, Orsay, France}
\newcommand*{\ORSAYindex}{23}
\affiliation{\ORSAY}
\newcommand*{\ITEP}{Institute of Theoretical and Experimental Physics, Moscow, 117259, Russia}
\newcommand*{\ITEPindex}{24}
\affiliation{\ITEP}
\newcommand*{\JMU}{James Madison University, Harrisonburg, Virginia 22807}
\newcommand*{\JMUindex}{25}
\affiliation{\JMU}
\newcommand*{\KNU}{Kyungpook National University, Daegu 41566, Republic of Korea}
\newcommand*{\KNUindex}{26}
\affiliation{\KNU}
\newcommand*{\MISS}{Mississippi State University, Mississippi State, MS 39762-5167}
\newcommand*{\MISSindex}{27}
\affiliation{\MISS}
\newcommand*{\UNH}{University of New Hampshire, Durham, New Hampshire 03824-3568}
\newcommand*{\UNHindex}{28}
\affiliation{\UNH}
\newcommand*{\NSU}{Norfolk State University, Norfolk, Virginia 23504}
\newcommand*{\NSUindex}{29}
\affiliation{\NSU}
\newcommand*{\OHIOU}{Ohio University, Athens, Ohio  45701}
\newcommand*{\OHIOUindex}{30}
\affiliation{\OHIOU}
\newcommand*{\ODU}{Old Dominion University, Norfolk, Virginia 23529}
\newcommand*{\ODUindex}{31}
\affiliation{\ODU}
\newcommand*{\RPI}{Rensselaer Polytechnic Institute, Troy, New York 12180-3590}
\newcommand*{\RPIindex}{32}
\affiliation{\RPI}
\newcommand*{\URICH}{University of Richmond, Richmond, Virginia 23173}
\newcommand*{\URICHindex}{33}
\affiliation{\URICH}
\newcommand*{\ROMAII}{Universita' di Roma Tor Vergata, 00133 Rome Italy}
\newcommand*{\ROMAIIindex}{34}
\affiliation{\ROMAII}
\newcommand*{\MSU}{Skobeltsyn Institute of Nuclear Physics, Lomonosov Moscow State University, 119234 Moscow, Russia}
\newcommand*{\MSUindex}{35}
\affiliation{\MSU}
\newcommand*{\SCAROLINA}{University of South Carolina, Columbia, South Carolina 29208}
\newcommand*{\SCAROLINAindex}{36}
\affiliation{\SCAROLINA}
\newcommand*{\TEMPLE}{Temple University,  Philadelphia, PA 19122 }
\newcommand*{\TEMPLEindex}{37}
\affiliation{\TEMPLE}
\newcommand*{\JLAB}{Thomas Jefferson National Accelerator Facility, Newport News, Virginia 23606}
\newcommand*{\JLABindex}{38}
\affiliation{\JLAB}
\newcommand*{\UTFSM}{Universidad T\'{e}cnica Federico Santa Mar\'{i}a, Casilla 110-V Valpara\'{i}so, Chile}
\newcommand*{\UTFSMindex}{39}
\affiliation{\UTFSM}
\newcommand*{\EDINBURGH}{Edinburgh University, Edinburgh EH9 3JZ, United Kingdom}
\newcommand*{\EDINBURGHindex}{40}
\affiliation{\EDINBURGH}
\newcommand*{\GLASGOW}{University of Glasgow, Glasgow G12 8QQ, United Kingdom}
\newcommand*{\GLASGOWindex}{41}
\affiliation{\GLASGOW}
\newcommand*{\VT}{Virginia Tech, Blacksburg, Virginia   24061-0435}
\newcommand*{\VTindex}{42}
\affiliation{\VT}
\newcommand*{\VIRGINIA}{University of Virginia, Charlottesville, Virginia 22901}
\newcommand*{\VIRGINIAindex}{43}
\affiliation{\VIRGINIA}
\newcommand*{\WM}{College of William and Mary, Williamsburg, Virginia 23187-8795}
\newcommand*{\WMindex}{44}
\affiliation{\WM}
\newcommand*{\YEREVAN}{Yerevan Physics Institute, 375036 Yerevan, Armenia}
\newcommand*{\YEREVANindex}{45}
\affiliation{\YEREVAN}

\newcommand*{\NOWCORNELL}{Cornell University, Ithaca, New York 14850}
\newcommand*{\NOWISU}{Idaho State University, Pocatello, Idaho 83209}
\newcommand*{\NOWLAMAR}{Lamar University, Beaumont, Texas, 77710}


\author {S.~Lombardo} 
\altaffiliation[Current address:]{\NOWCORNELL}
\affiliation{\INDIANA}
\author {M.~Battaglieri} 
\affiliation{\INFNGE}
\author {A.~Celentano} 
\affiliation{\INFNGE}
\author {A.~D'Angelo} 
\affiliation{\INFNRO}
\affiliation{\ROMAII}
\author {R.~De~Vita} 
\affiliation{\INFNGE}
\author {A.~Filippi} 
\affiliation{\INFNTUR}
\author {D.I.~Glazier} 
\affiliation{\GLASGOW}
\author {S.~M.~Hughes} 
\affiliation{\EDINBURGH}
\author {V.~Mathieu}
\affiliation{\INDIANA}
\affiliation{\IND}
\affiliation{\JLAB}
\author {A.~Rizzo} 
\affiliation{\INFNRO}
\affiliation{\ROMAII}
% ...

\collaboration{The CLAS Collaboration}
\noaffiliation


\begin{abstract}
The exclusive reaction $\gamma p \to p K^+ K^-$  was studied 
in the photon energy range $3.0 - 3.8 \mbox{ GeV}$ and momentum transfer range  $0.6<-t<1.3 \mbox{ GeV}^2$. Data were collected with the  CLAS detector at the Thomas Jefferson National Accelerator Facility. In this kinematic range
the integrated luminosity was approximately 20 pb$^{-1}$.
The  reaction was isolated by detecting the $K^+$ and the proton in CLAS, and reconstructing the  $K^-$ via the missing-mass technique. Moments of the di-kaon decay angular distributions
were extracted from the experimental data.
Besides  the dominant  contribution of the $\phi$ meson in the $P$-wave,  evidence for $S-P$ interference  was found.
The differential production cross sections $d\sigma/dt$ for individual waves  in the mass range of  the $\phi$ resonance were extracted and compared to predictions of a Regge-inspired model.  
This is the first time the $t$-dependent cross section of the $S$-wave contribution to the elastic $K^+K^-$ photoproduction has been measured.
\end{abstract}

\maketitle

\section{\label{sec:intro}Introduction}
 Data on light quark mesons comes mainly 
  from hadron induced reactions, {\it e.g.} by using $K$, $\pi$, $p$ or $\bar p$ beams and, more recently, from decays of heavy mesons. 
Up to now, only a few studies of the light meson spectrum were attempted with electromagnetic probes and, in particular, with real photons.
  The main reason for
 this is the relatively small production 
cross sections compared to hadronic reactions. 
However, this situation is changing,   thanks to the recent advances in producing high-intensity and high-quality tagged, polarized photon beams.  At  lower energies, {\it e.g.} near single meson production thresholds, high quality data have been accumulated  by the CB-ELSA~\cite{elsa}  and CB-MAMI~\cite{Ostrick:2016eig} experiments, while at higher energies, photoproduction data have come from the CLAS~\cite{Ireland:2017ksn} experiment at Jefferson Lab. Moreover, two new programs,  GLUEX~\cite{Patsyuk:2017imk} and MesonEx  ~\cite{mesonx} have just been launched in the same laboratory. 
 A typical meson photoproduction data set from past experiments in the energy range below 20 GeV, typical for meson spectroscopy, has tens of thousands of events, and only a few topologies have been studied~\cite{Ballam73,Aston80,Fries78}. 
For comparison, the data samples from the $g11$ run at CLAS used here, exceed the existing sets in many channels by at least an order of magnitude, and several reconstructed topologies are available for a comprehensive study~\cite{Battaglieri:2012zz}.

Specifically, two-pseudoscalar meson photoproduction (two-pion and two-kaon) offers the possibility of investigating various aspects of the light meson resonance spectrum. 
Two-pion is the main decay mode of the  lowest isoscalar-tensor, the  $f_2(1270)$ resonance, and it is 
the only known hadronic decay mode of the lowest isovector-vector resonance, the $\rho(770)$. 
The two-kaon channel is the main decay mode of the isoscalar-vector $\phi(1020)$ and a possible sub-threshold decay of the isoscalar-scalar $f_0(980)$ and the isovector-scalar $a_0(980)$.
 Both the two pion and two kaon decay modes couple to the isoscalar-scalar channel, which contains the  $f_0(500)$ and  $f_0(980)$ resonances ~\cite{Pelaez:2015qba} and a few more resonances with masses above 1 GeV that are not yet well understood.   For example, the $f_0(500)$  meson, which is now well established~\cite{Caprini:2005zr,Kaminski:2006qe,Kaminski:2006yv}, but does not fit the naive quark model classification. The $f_0(980)$ is similarly difficult to classify and its composition is affected by proximity to the $K{\bar K}$ threshold. 
These states have been the subject of extensive investigations~\cite{Dai:2014zta,Briceno:2017qmb} since their 
observation in photon induced reactions can provide insights into their internal structure.\\
In this paper we present results of the analysis of
 $K^+K^-$ photoproduction in the photon energy range $3.0 - 3.8 \mbox{  GeV}$ and momentum transfer 
squared $-t$  between $0.6 \mbox{ GeV}^2$ and $1.3 \mbox{ GeV}^2$, where
the di-kaon effective mass $M_{K^+K^-}$ varies from 0.990 to 1.075 GeV. 
We have focused on  this mass region because it is dominated by the production of the $\phi(1020)$ resonance that decays to the two kaons in  the $P$-wave, and thus a partial wave analysis based on the lower ($S$ and $P$) waves efficiently describes it.  To describe the higher mass region would require a higher number of partial waves, and is not included in this study.

Angular distributions of photoproduced mesons and related observables, such as the spherical harmonic moments and the spin density matrix elements, are the most effective tools
for studying individual partial waves. For example, interference between the $S$-wave and the dominant $P$-wave
was first discovered in the moment analysis of $K^+K^-$ photoproduction on hydrogen in the experiments performed at DESY~\cite{Behrend}  and Daresbury~\cite{Barber}.
In this work we applied the same methodology used in the 
analysis of two pion photoproduction to the same data set~\cite{f0-clas,2pi-clas} and we refer the reader to those  works for a detailed description of the analysis procedure.

This paper is organized as follows. In the next section we give a summary of the experimental setup and data analysis.   Extraction of  the angular moments of the 
two-kaon system is described in Section ~\ref{sec:mom}. The  
fit of a phenomenological model to the extracted moments is described in 
 Section~\ref{sec:disp}, where we also present results of the partial wave analysis, including the extracted differential cross sections for each partial wave, and a physics interpretation. A summary of the results is given in  Section~\ref{sec:sum}. 

\section{\label{sec:exp} Experimental procedures and data analysis}
\subsection{The photon beam and the target}
The measurement   was performed with the CLAS detector~\cite{CLAS} 
in  Hall B at Jefferson Lab with a bremsstrahlung 
photon beam  produced by a continuous 60 nA electron beam of energy  $E_0$ = 4.02 GeV  
impinging on a gold foil of thickness $8 \times 10^{-5}$ radiation lengths.
A bremsstrahlung tagging system~\cite{SO99} with a photon energy resolution of 0.1$\%$ $E_0$ 
was used to tag photons in the energy range from 1.6 GeV
to a maximum energy of 3.8 GeV. In this analysis only the high-energy 
part of the photon spectrum, ranging from 3.0 to 3.8 GeV, was used. The
$e^+$ $e^-$ pairs produced by  interactions of the  photon beam on an additional thin gold foil were used
to continuously monitor the photon flux  during the experiment. Absolute normalization was obtained by comparing 
the $e^+$ $e^-$ pair rate with the photon flux measured by a total absorption lead-glass counter in dedicated
low-intensity runs.
The  energy calibration of the Hall-B tagger system
was performed both by a direct measurement of the $e^+e^-$ pairs produced by the incoming photons
and by applying  an over-constrained kinematic fit  to the  reaction $\gamma p \to p \pi^+ \pi^-$, where all particles
in the final state were detected in CLAS~\cite{tag-abs_cal}.
The quality of the calibrations was checked by 
looking at  the mass  of known particles, as well as their dependence on  other kinematic variables 
(photon energy, detected particle momenta and angles).

The target cell, a Mylar  cylinder  4 cm in diameter and 40-cm long, was filled by liquid hydrogen at 20.4 K.
The luminosity was obtained as the product of the target density, 
target length and the incoming photon flux 
corrected for data-acquisition dead time.
The overall   systematic uncertainty on the run  luminosity was estimated to be approximately  10$\%$,
dominated by the uncertainty of the photon flux normalisation~\cite{devita}.

\subsection{The CLAS detector}
Outgoing  hadrons were detected in the CLAS  spectrometer.
Momentum information for charged particles was obtained via tracking
through three regions of multi-wire drift chambers~\cite{DC} within  a toroidal magnetic 
field ($\sim 1.25$ T) generated by six superconducting coils. 
The polarity of the field was set to bend the positive particles away from the beam line  into the acceptance 
of the  detector.
Time-of-flight scintillators (TOF) were used for charged hadron
identification~\cite{Sm99}. 
The interaction time between the incoming photon and the target
was measured by  the start counter (ST)~\cite{ST}. This   was
made of 24 strips of 2.2 mm thick plastic scintillator surrounding the hydrogen cell
with a single-ended PMT-based read-out. 
The average time resolution of the ST strips was $\sim$300~ps.

The CLAS momentum resolution, $\sigma_p/p$, ranged from 0.5 to 1.0\%, depending on
the kinematics. 
The detector geometrical acceptance for each positive particle in the 
relevant kinematic region was about 40\%. It was somewhat less for low-energy negative 
hadrons, which could be lost at  forward angles because
their paths were bent toward the beam line and out of the acceptance
by the toroidal field.
Coincidences between the photon tagger and the CLAS detector triggered 
the recording of the events. The trigger in CLAS  required
a coincidence between the TOF and the ST 
in at least two sectors, in order to  select
reactions with at least two charged particles in the final state.
A total integrated luminosity of 70 pb$^{-1}$ ($\sim20$ pb$^{-1}$ in the range 3.0$<E_\gamma<$3.8 GeV)
was accumulated in  50 days of data taking  in 2004.

\subsection{Data analysis and reaction identification}\label{ssec:reac_id}
The raw data were passed through the standard CLAS reconstruction software to determine the four-momenta of the detected particles.
In this phase of the analysis, corrections were applied to account for the energy loss of charged particles in the target and 
surrounding  materials, misalignments of the  drift chamber positions, and 
uncertainties in the value of the toroidal magnetic field.\\
\begin{figure}
\center
\includegraphics[width=0.5\textwidth]{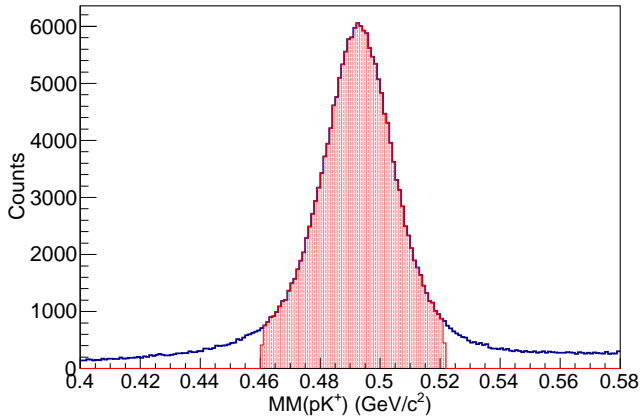} 
\caption{Missing mass of the reconstructed $K^-$ for the reaction $\gamma p \to p K^+ K^-$. Only events in the shaded area were used in the analysis. \label{fig:missmass}}
\end{figure}
The reaction $\gamma p \to p K^+ K^-$ was isolated by detecting the  proton and the $K^+$ in the CLAS spectrometer,
while the  $K^-$ was reconstructed from the four-momenta of the detected particles by using the missing-mass technique.
A combination of drift chambers and TOF information allowed for the identification of  the kaon band in the $\beta$ vs. $p$ plane for positive charged particles. More details, as well as  the  resulting $K^+$ missing mass  spectrum for the reaction $\gamma p \to K^+ X$ can be found in 
Ref.~\cite{devita}.
The exclusivity of the reaction was ensured by retaining events  within 3$\sigma$ around the missing $K^-$ peak (492 MeV $\pm$ 30 MeV). This cut
kept the contamination from pion misidentification and multi-kaon  background to a minimum ($\sim$7\%) for events in the di-kaon mass range of interest for this analysis (0.990 GeV $<M_{K^+K^-}<$ 1.075 GeV). 
\ Figure~\ref{fig:missmass} shows the $K^-$ missing mass.
The  background below the kaon peak appears as a smooth contribution 
to the $K^+K^-$ invariant mass that can be accounted for by fitting and subtracting a polynomial function.
Since the focus of the paper is about the interference of the narrow $P$-wave (the $\phi$ meson) with the  $S$-wave,  the experimental background, as well as the  projection of high mass hyperons populating the $pK^+$ mass spectrum, 
enters in the $K^+K^-$ mass as a smooth  incoherent contribution that does not affect the results.

\begin{figure}
\center
\includegraphics[width=0.5\textwidth]{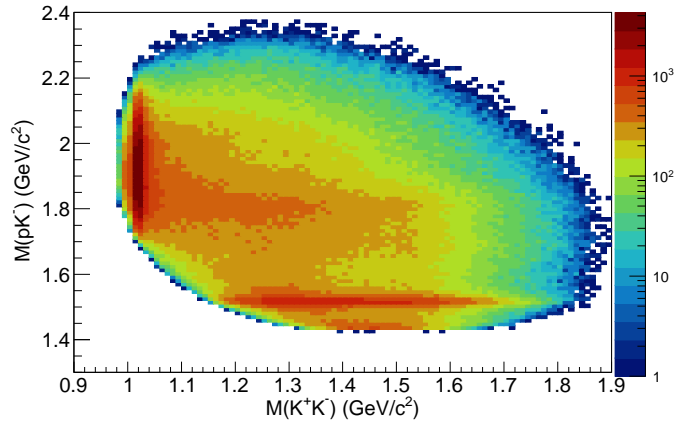} 
\caption{Invariant mass of the $pK^+$ system vs. invariant mass of the $K^+K^-$
         system. The $\phi$ meson shows up as a narrow vertical band peaked around 1~GeV, while the $\Lambda(1520)$ is visible as a horizontal band around 1.5 GeV. \label{fig:dalitzplot}}
\end{figure}

To  cut out  edge regions in the  detector acceptance, only events within a {\it fiducial phase space} volume were retained in this analysis.
In the laboratory reference system, cuts were defined for
the minimum hadron momentum ($p_{p}>0.32$~GeV/c and $p_{K^+}>0.125$~GeV/c), and the minimum  
angles ($\theta_{p} >10^\circ$ and $\theta_{K^+} >5^\circ$). 
The  fiducial cuts  were defined comparing in detail the experimental data distributions with the results of  the detector  simulation.
The minimum momentum cuts were tuned for different hadrons to take into account the energy loss as the particles pass through the target and the detector.\\ 
After all cuts, 0.2M  events were identified as produced in the exclusive reaction   $\gamma p \to p K^+ (K^-)$.
The other event topologies that required the $K^-$ to be detected
were not used since, in the kinematics of interest for this analysis ($-t<1.3$~GeV$^2$), 
the collected data were about one  order of magnitude less due to the reduced detector acceptance for the inbending $K^-$.
Figure ~\ref{fig:dalitzplot} shows the invariant mass spectra of $pK^-$ and $K^+K^-$ using the reconstructed $K^-$  four-momentum.\\
The $\phi(1020)$ dominates the $K^+ K^-$ spectrum and  the $\Lambda(1520)$ peak is  visible in the mass spectrum of the  $p K^-$
invariant mass. No overlap between  the $\Lambda(1520)$ peak  and the $K^+ K^-$ spectrum occurs for 
$M_{K^+K^-}<1.25$ GeV. Nevertheless, a sharp cut for $M_{pK^-}<1.6$ GeV was applied to avoid any contamination in the meson spectrum from the    $\Lambda(1520)$.
A hint of excited $\Lambda$ states is visible in the bi-dimensional distribution  but their contribution to the $K^+K^-$ spectrum is very small and tends to be smooth when all hyperon states are integrated over.

\section{\label{sec:mom} Moments of the di-kaon angular distributions}\label{par:fin_results} 
In this section we consider the analysis of spherical harmonic moments, 
$ \langle Y_{\l\m} \rangle =  \langle Y_{\l\m} \rangle(E_\gamma,t,M_{K^+K^-})$, of the  di-kaon angular distribution defined as,
\begin{equation}\label{eq:mom}
\langle Y_{\l\m} \rangle = \sqrt{4\pi} \int d\Omega_K  {{d\sigma} \over {dt\;dM_{K^+K^-} \;d\Omega_K}} Y_{\l\m}(\Omega_K),
\end{equation} 
where $d\sigma$ is the four-fold differential cross section at fixed 
 photon energy $E_\gamma$. Here $t$ is the momentum transfer squared between the target and the recoil proton,  
  $M_{K^+K^-}$ is the di-kaon invariant mass and $Y_{\l\m}$ are
spherical harmonics. The spherical angle 
$\Omega_K = (\theta_K , \phi_K)$ corresponds to the 
 direction of flight of the  $K^+$  
in the $K^+K^-$ helicity rest frame. This is the rest frame of the $K^+  K^-$ pair, with the $y$-axis perpendicular to the production plane and the $z$-axis pointing in the opposite direction of the recoil nucleon  momentum. 
 In  equation \eqref{eq:mom}  the normalization has been chosen such that 
  the $\langle Y_{00}\rangle$ moment is equal to the di-kaon production  differential  cross section $d\sigma/dt\;dM_{K^+K^-}$. 

There are several advantages in using moments of the angular distribution compared to a direct partial wave analysis. 
Moments  can  be  expressed as bi-linear in terms of the partial waves and, 
depending on the particular combination of $L$ and $M$,  show specific sensitivity to  a particular subset of them.
In addition, they can be directly and
unambiguously derived from the data, allowing for a quantitative comparison to the same observables calculated in specific theoretical models. Since partial wave analysis has either intrinsic mathematical ambiguities or is model dependent, it is important to extract physical observables like moments before proceeding with a model dependent analysis~\cite{Chung:1997qd}.

 The moments were extracted using two separate methods, both expanding  in a model-independent set of basis functions, which were compared to the data by maximizing a likelihood function. The  first of these two methods (M1) parametrized the angular distributions in terms of moments directly, while the second method (M2) used spherical harmonic partial wave amplitudes.
 The approximations in these two methods are dependent on the basis and
on their truncation.
As a check of systematics we also applied two further methods:  we first binned the data and Monte Carlo simulations in all kinematical variables and divided the data by acceptance to obtain the expected angular distributions; the second used linear algebra techniques to set up an over-determined system of equations for the moments. They provided consistent results but  were not as stable or reliable as the maximum likelihood methods M1 and M2 and were  not included in the  final determination of the experimental moments.
Detailed systematic studies using both Monte Carlo and data were performed to test the stability of the results for the different methods. A summary of these studies is reported in Appendix~\ref{appendix:syst}.
Full details regarding the procedure adopted for  the moment extractions  are reported in~\cite{2pi-clas,sal-note}.

\subsection{Detector efficiency}
The CLAS detection efficiency for the reaction  $\gamma p \to p K^+K^-$ was obtained by means of detailed Monte Carlo 
simulations, which included knowledge of the full detector geometry and a realistic response to traversing particles. Events were generated 
according  to three-particle phase space with a  bremsstrahlung photon energy spectrum.
A total  of 96 M  events were generated in the energy range  3.0 GeV $< E_\gamma <$  3.8 GeV  and covered
the allowed kinematic  range in $-t$ and $M_{K^+K^-}$. About 19 M  events were reconstructed 
in the $M_{K^+K^-}$ and $-t$ ranges of interest
(0.990 GeV $< M_K<$ 1.365 GeV, 0.6 GeV$^2< -t < $ 1.3 GeV$^2$). 
This corresponds to more than 400 times the  statistics collected in the experiment, thereby introducing a negligible 
statistical uncertainty with respect to the statistical fluctuations  of the data. 

\subsection{Extraction of the moments via likelihood fit of experimental data}
The extraction of the moments, $\langle Y_{LM}\rangle$, was performed using the extended maximum likelihood method. As stated above, the expected theoretical  yield was parametrized in terms of appropriate functions, amplitudes in one case and  moments in the other. The theoretical expectation, after correction  for acceptance, was  compared to the experimental yield. The likelihood is then given by,
\begin{equation}
{\cal L}\sim\frac{\overline{n}_{Y_{LM}}^{n}}{n!}e^{-\overline{n}_{Y_{LM}}}\Pi_{a=1}^{n}\left[\frac{\eta(\tau_{a})I(\tau_{a},\langle Y_{LM}\rangle)}{\bar{n(\langle Y_{LM}\rangle)}}\right].
\end{equation} 

Here  $a$ represents a data event, $n $ is the number of data events in a given $(E_\gamma,t,M_{K^+K^-})$ bin ({\it i.e.}
the fit is done independently in each bin), $\tau_a$ represents  the set of kinematical variables of the $a^{th}$ event (here the two kaon decay angles),
 $\eta(\tau_a)$ is the corresponding acceptance derived by Monte Carlo simulations  and $I(\tau_a)$ is the theoretical 
function representing the expected event distribution.
The measure $d\tau$ includes the phase space factor and the likelihood function is normalized to the expected number of events in the bin
\begin{equation} 
\overline{n}_{Y_{LM}}=\int d\tau\eta(\tau)I(\tau,\langle Y_{LM}\rangle).
\end{equation} 
This normalization integral was performed by Monte-Carlo integration over the reconstructed simulated events.
The parameters were extracted by minimizing a function of the form,
\begin{equation} 
-2\ln{\cal L}\propto-2\sum_{a=1}^{n}I(\tau_{a},\langle Y_{LM}\rangle)+2\;\overline{n}_{Y_{LM}}.
\end{equation} 
The advantage of this approach lies in avoiding binning the data and the large uncertainties related to the corrections
in  regions of CLAS with vanishing efficiencies. 

Comparison of the results of the two different extraction methods
allows one to estimate the systematic uncertainty related to the procedure.
A detailed description of the two approaches is reported in Ref.~\cite{2pi-clas}.

\begin{figure}
\includegraphics[width=0.42\textwidth]{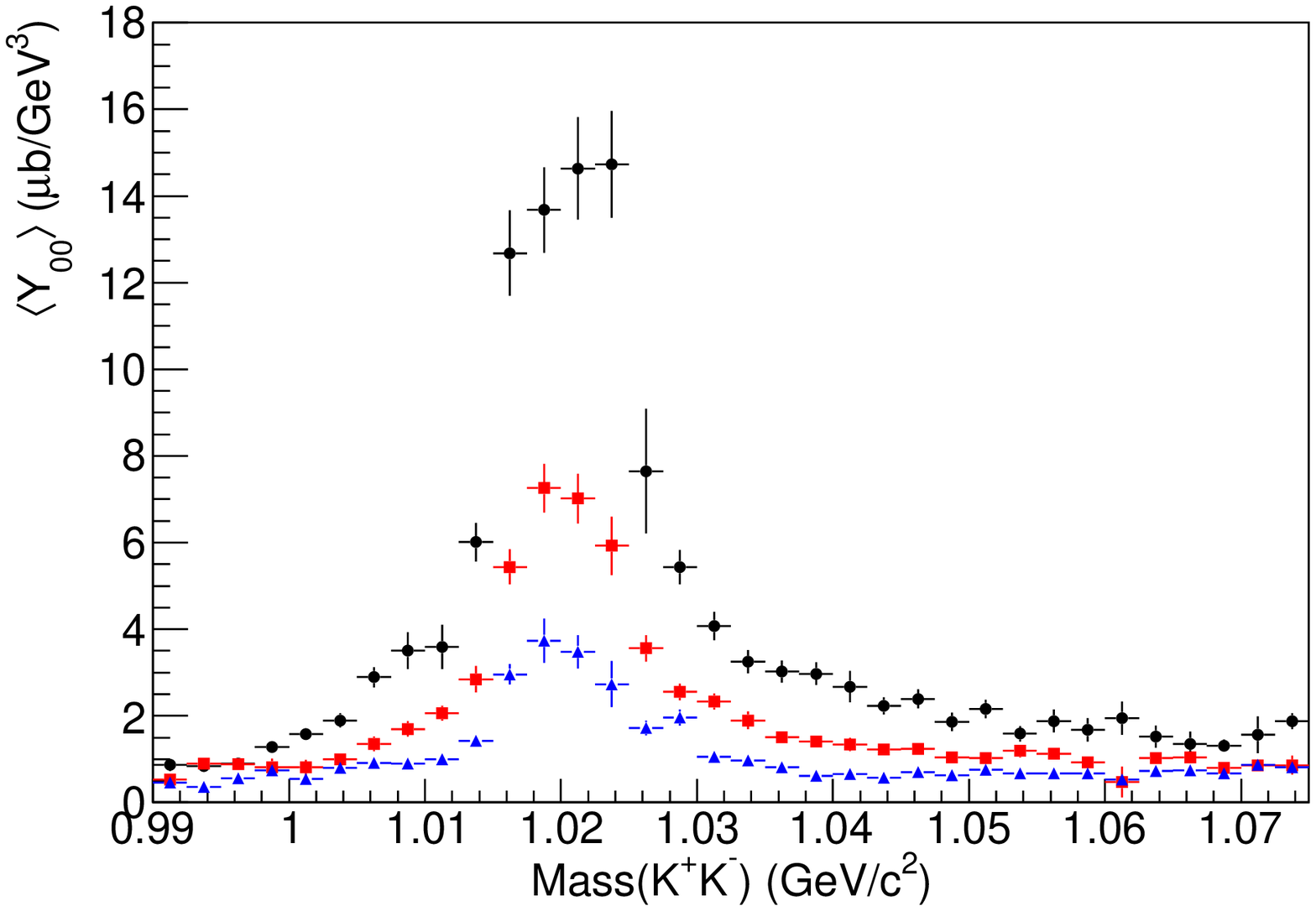} 
\includegraphics[width=0.42\textwidth]{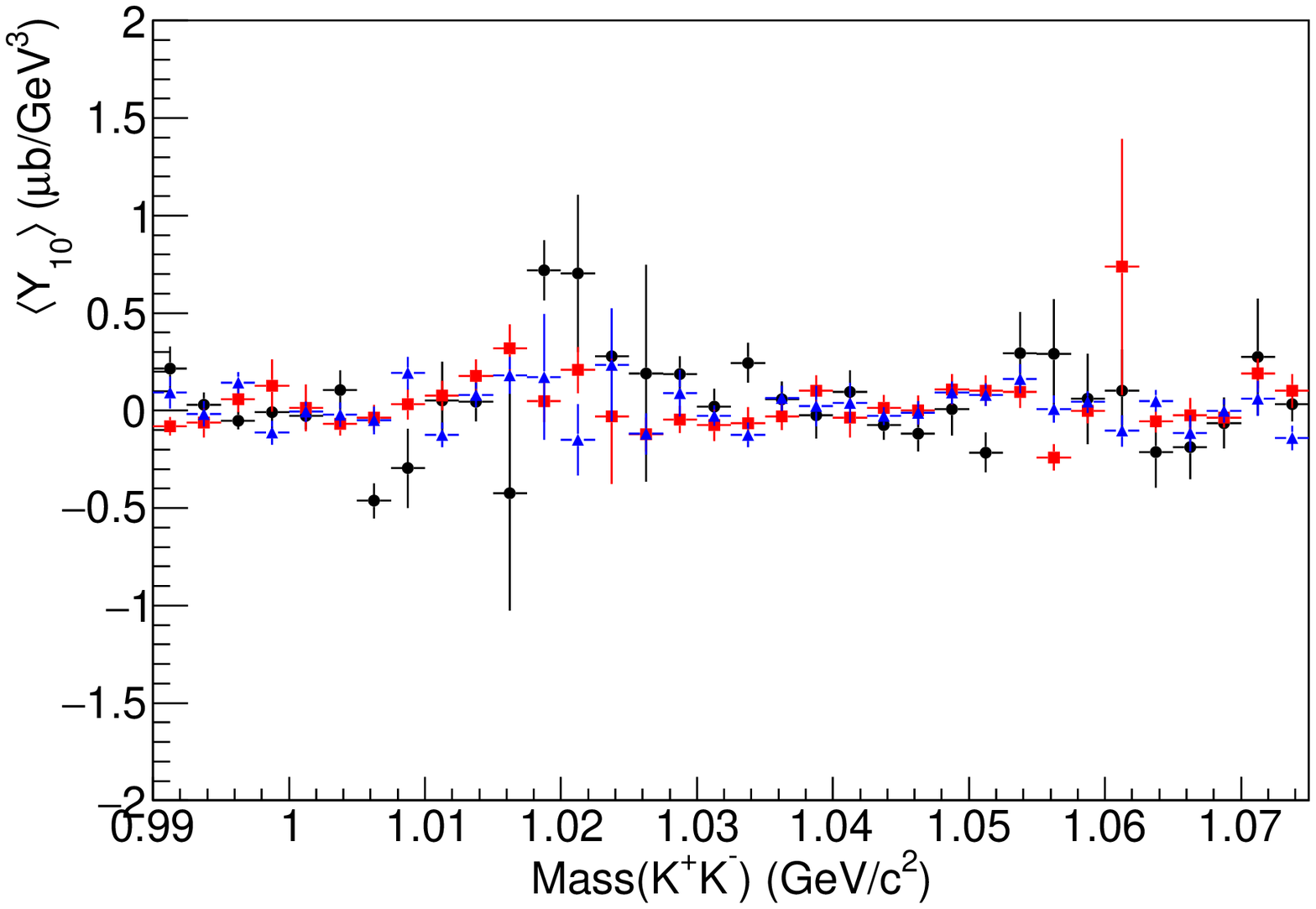} 
\includegraphics[width=0.42\textwidth]{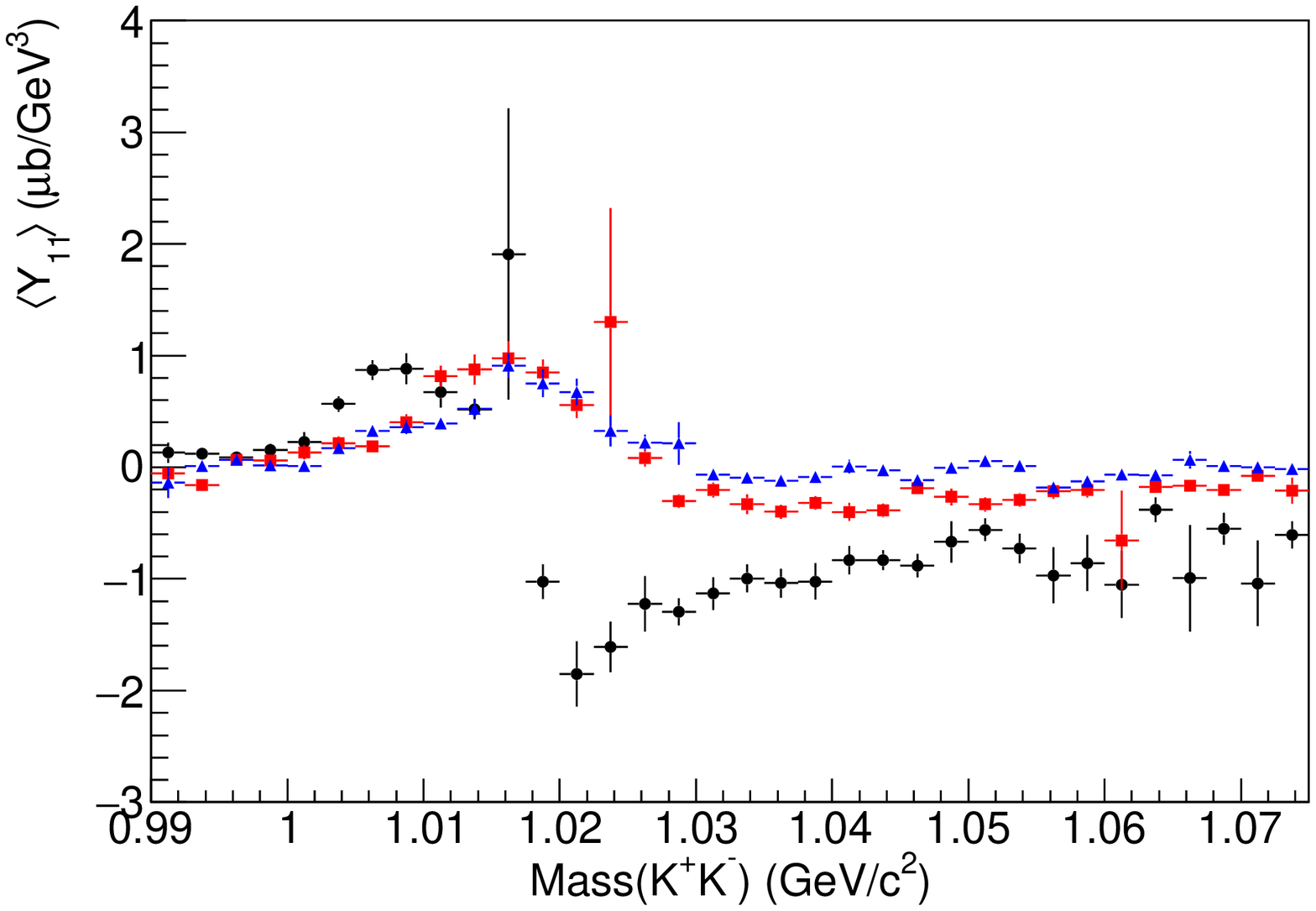} 
\includegraphics[width=0.42\textwidth]{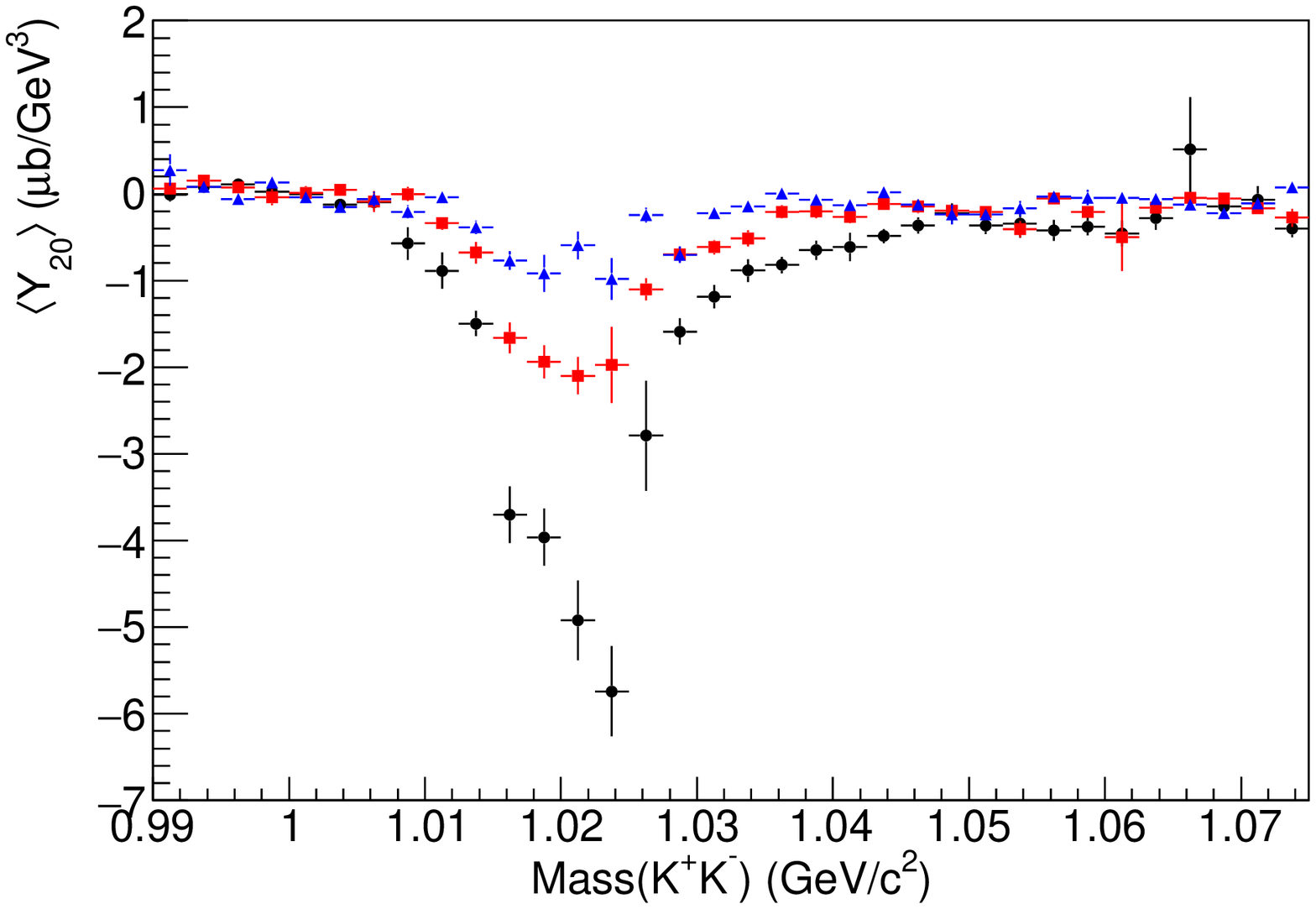} 
\caption[]{Moments of the di-kaon angular distributions  for $3.0 <E_\gamma< 3.8 \mbox{ GeV}$  
and $-t=0.45\pm0.05 \mbox{ GeV}^2$ (black), $-t=0.65\pm0.05 \mbox{ GeV}^2$ (red) and $-t=0.95\pm0.05 \mbox{ GeV}^2$ (blue). The error bars include both statistical and systematic uncertainties as explained in the text.}
\label{fig:final-2}
\end{figure}
\begin{figure}
\includegraphics[width=0.42\textwidth]{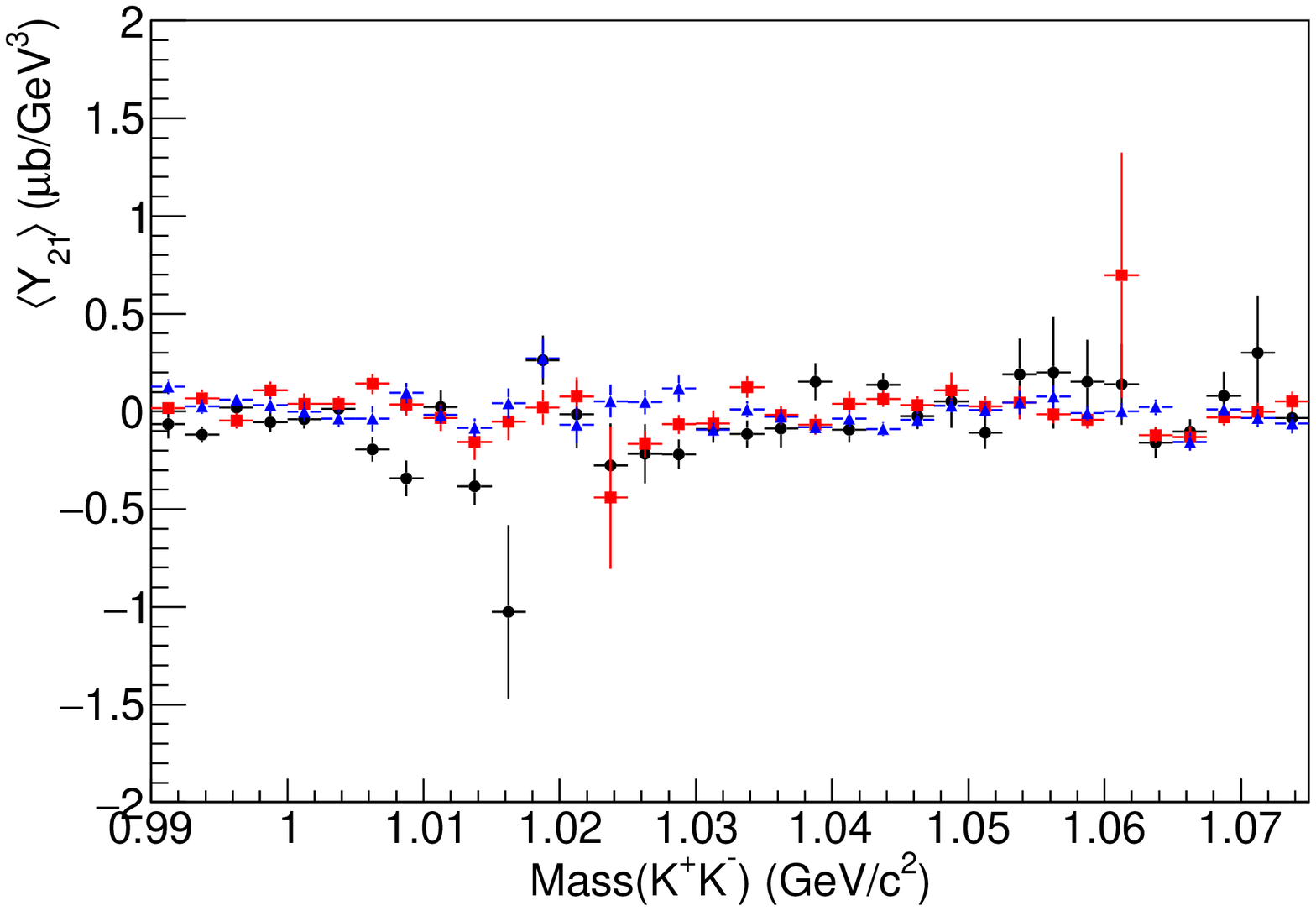} 
\includegraphics[width=0.42\textwidth]{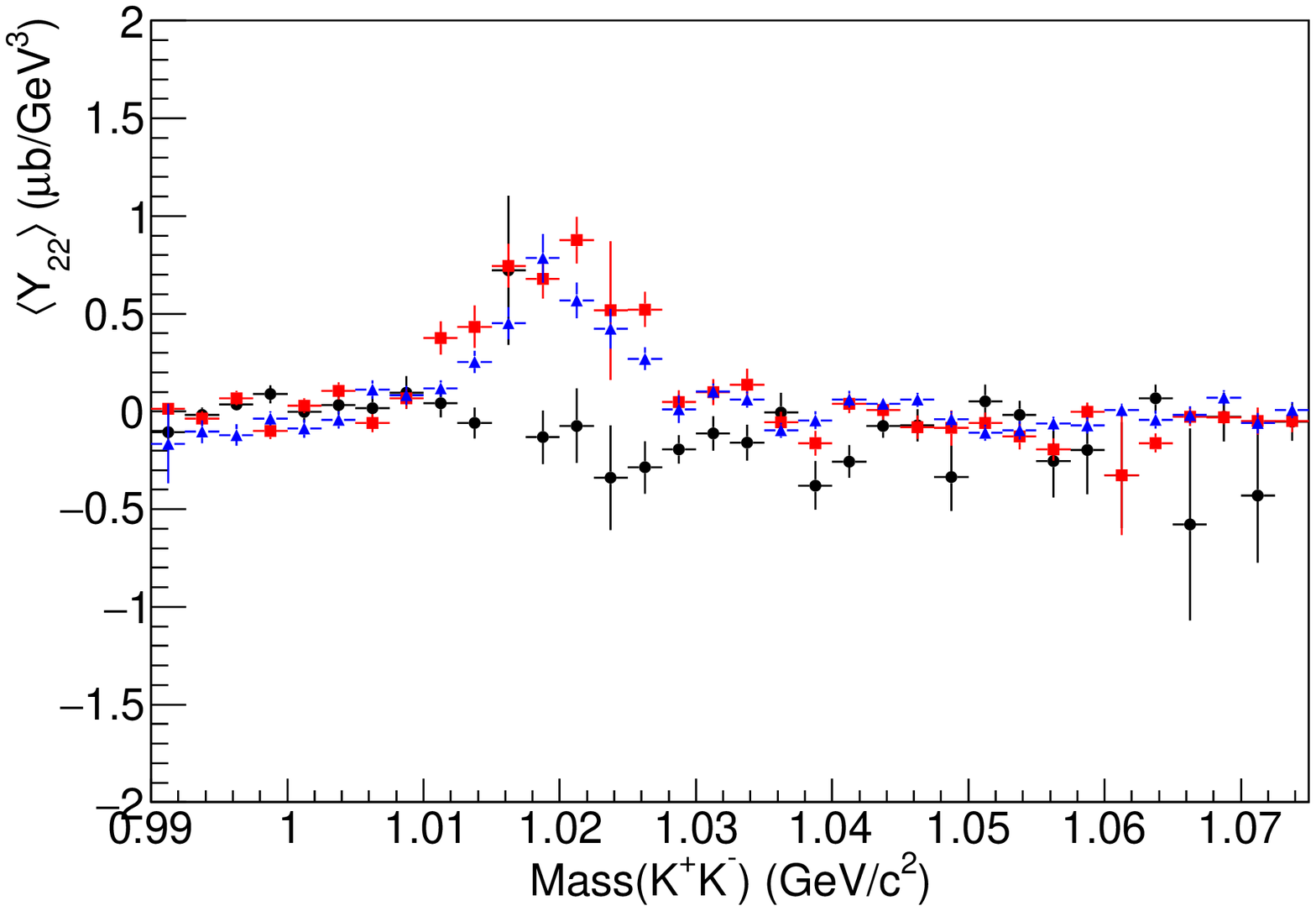} 
\includegraphics[width=0.42\textwidth]{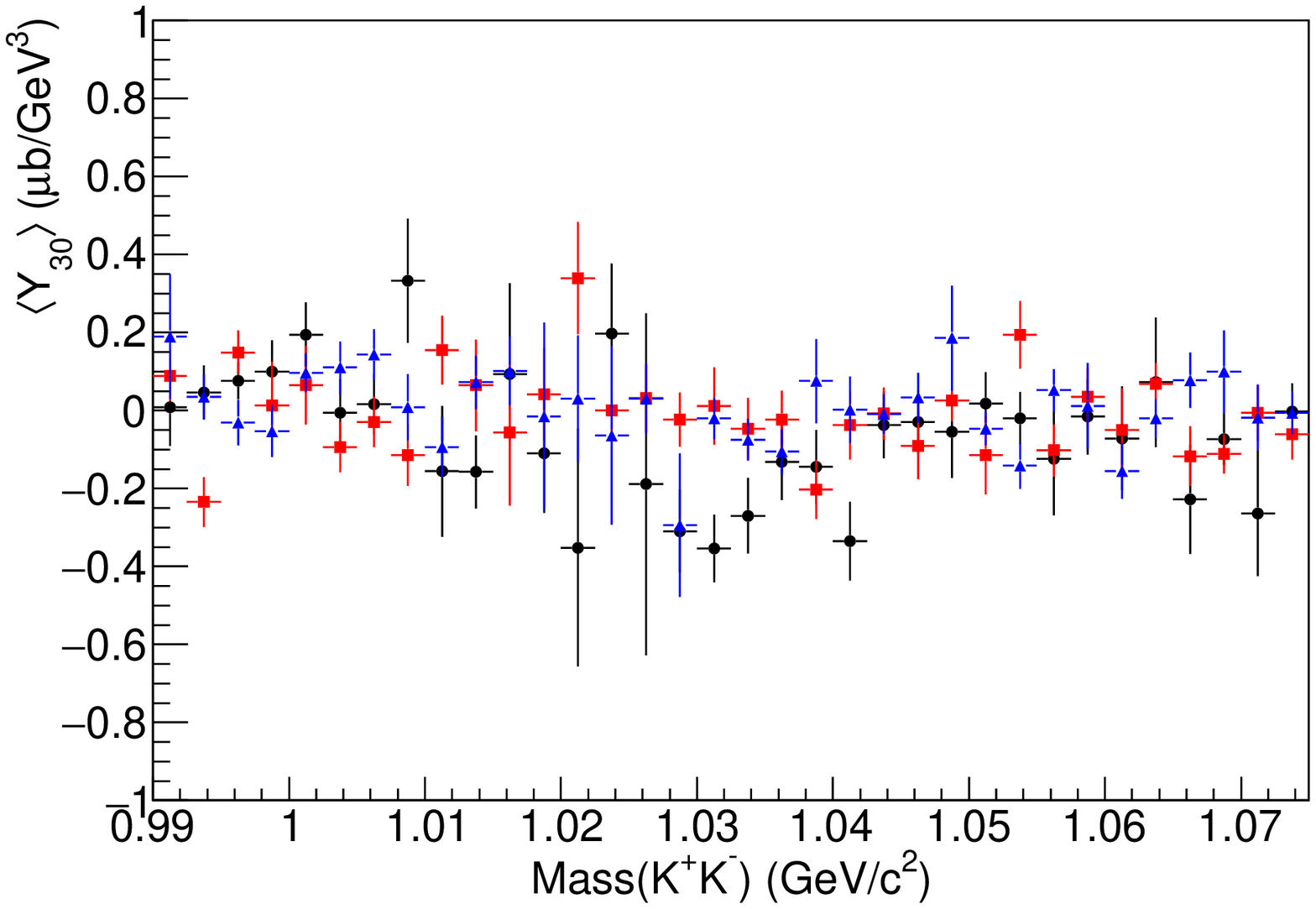} 
\includegraphics[width=0.42\textwidth]{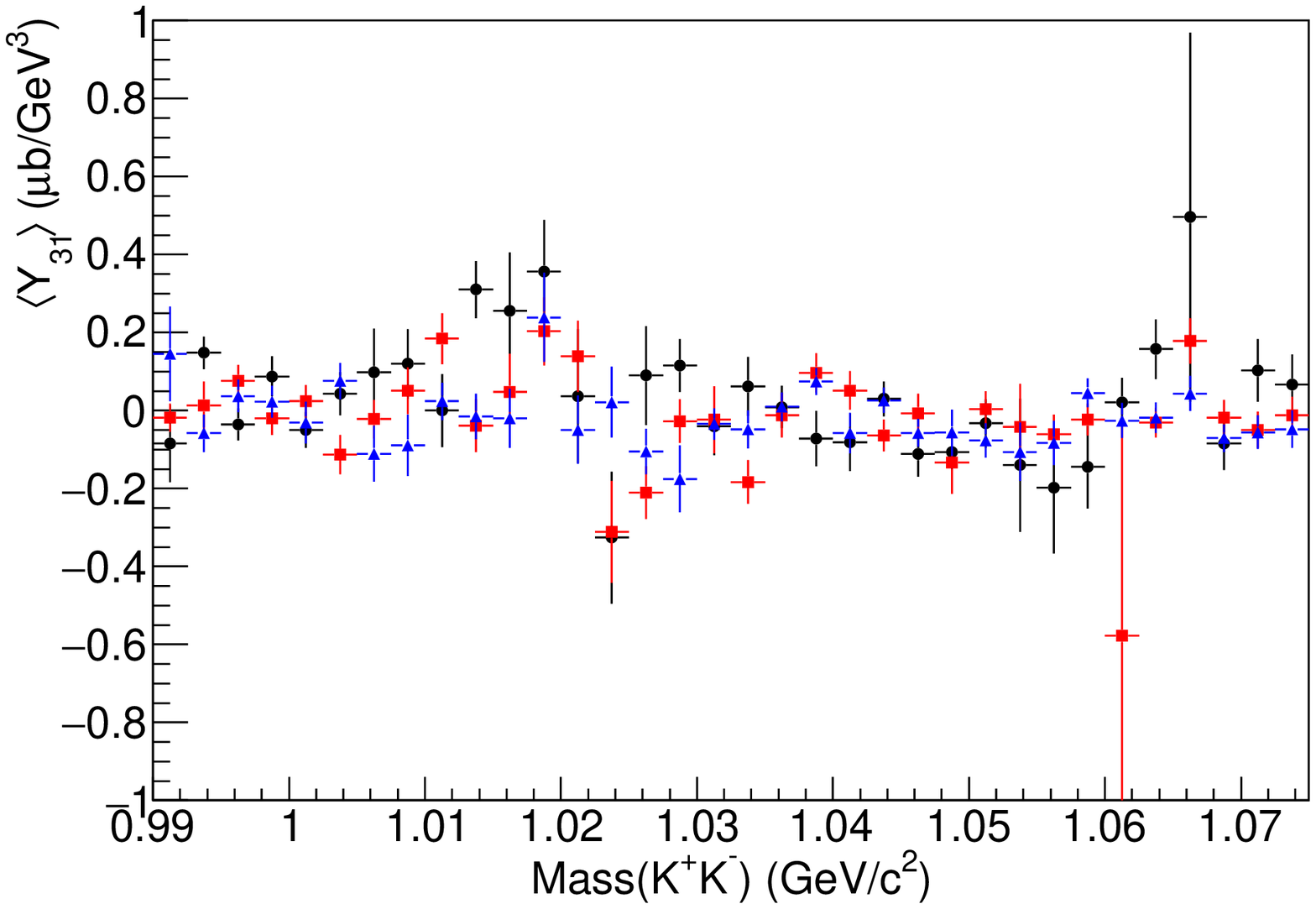} 
\caption[]{Moments of the di-kaon angular distributions  for $3.0 <E_\gamma< 3.8 \mbox{ GeV}$  
and $-t=0.45\pm0.05 \mbox{ GeV}^2$ (black), $-t=0.65\pm0.05 \mbox{ GeV}^2$ (red) and $-t=0.95\pm0.05 \mbox{ GeV}^2$ (blue). The error bars include both statistical and systematic uncertainties as explained in the text. }
\label{fig:final-3}
\end{figure}
\begin{figure}
\includegraphics[width=0.42\textwidth]{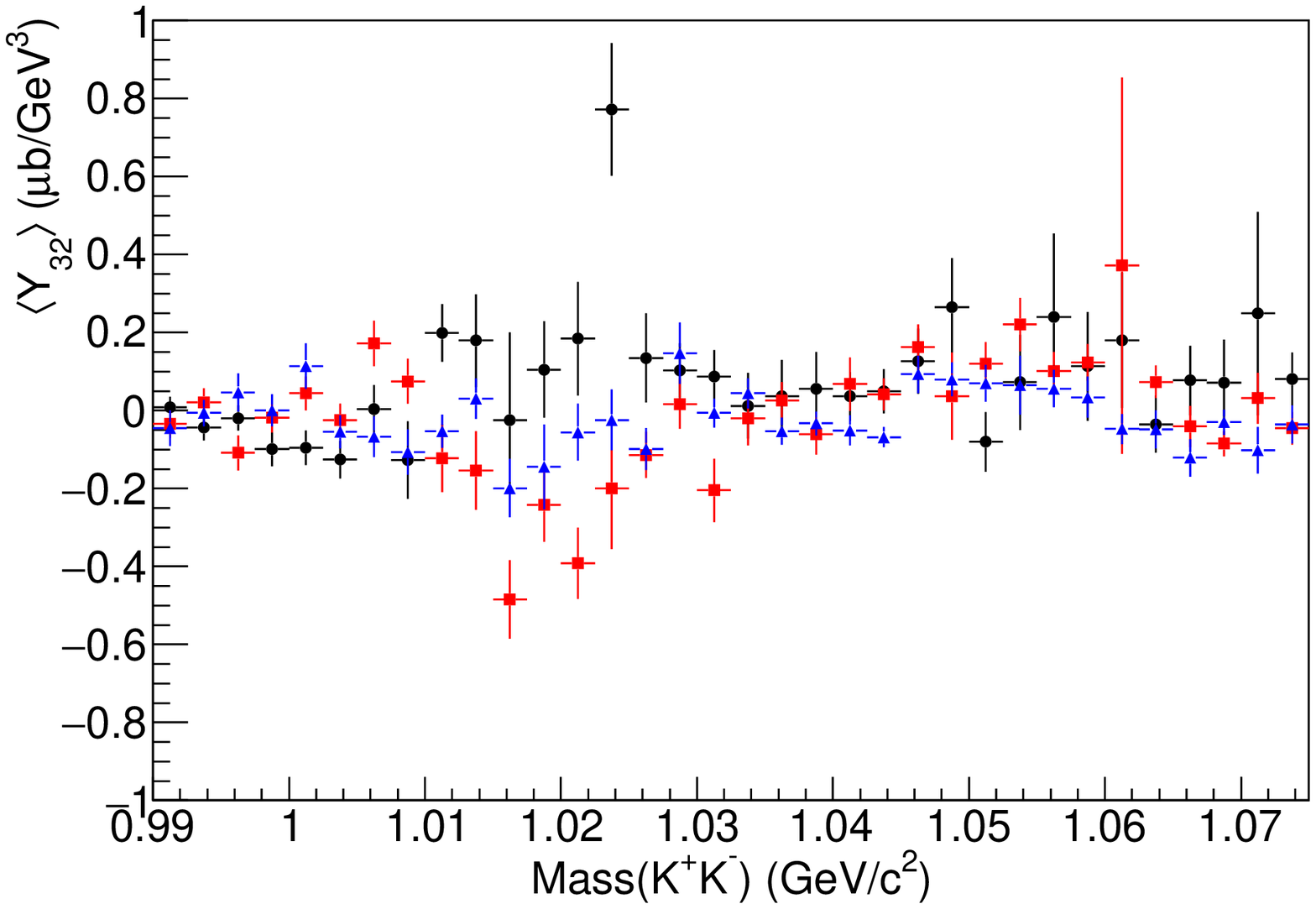} 
\includegraphics[width=0.42\textwidth]{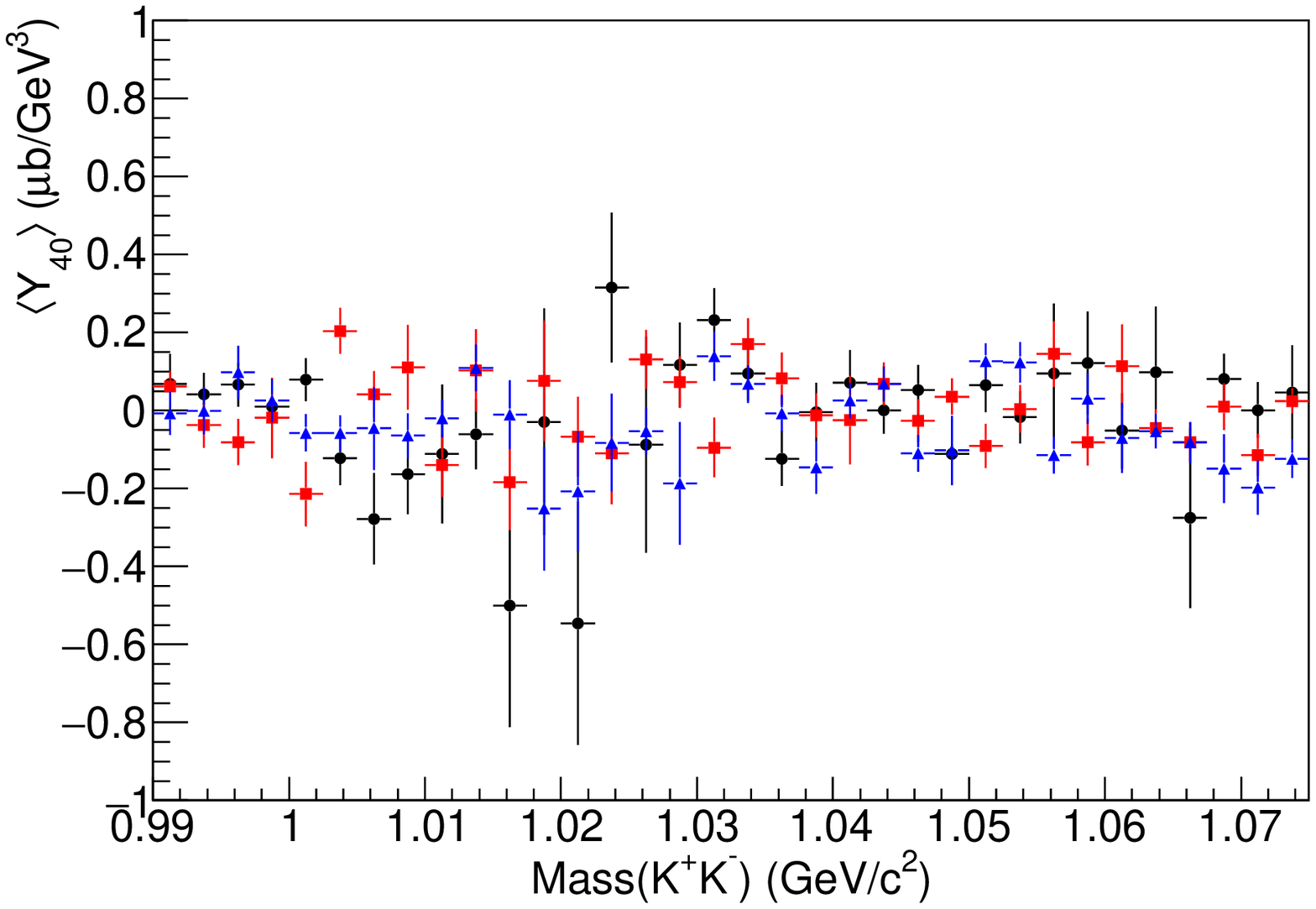} 
\includegraphics[width=0.42\textwidth]{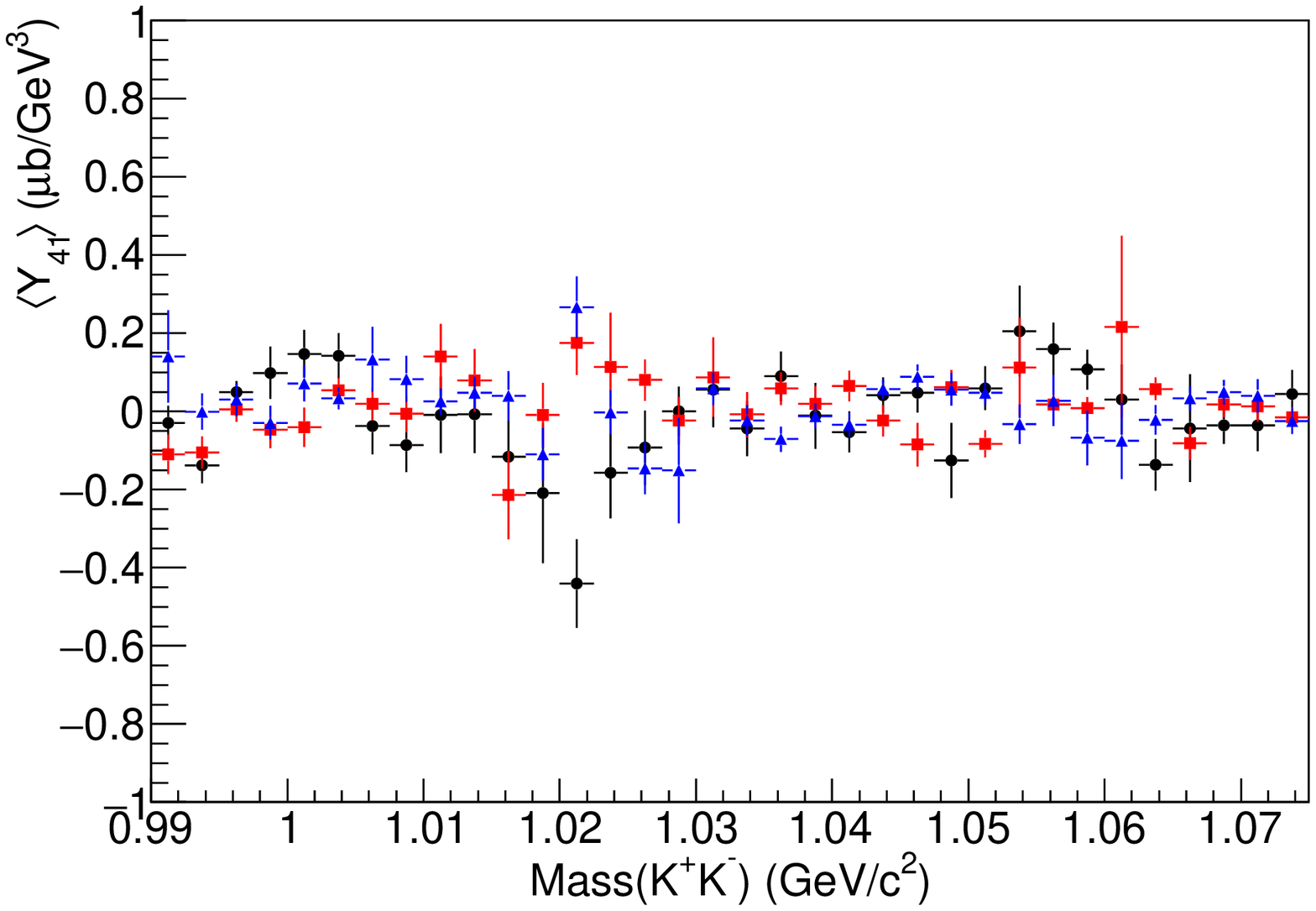} 
\includegraphics[width=0.42\textwidth]{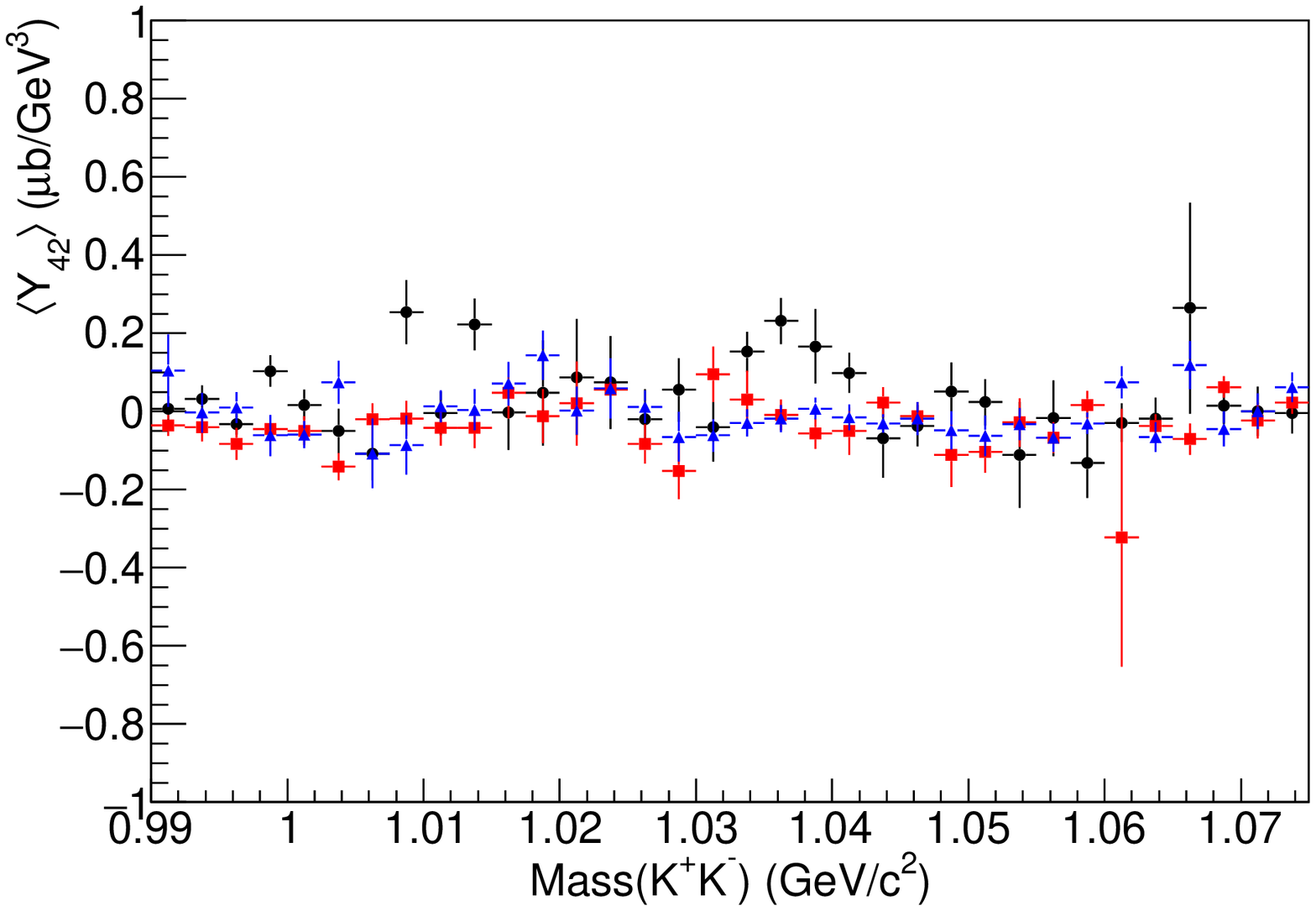} 
\caption[]{Moments of the di-kaon angular distributions  for $3.0 <E_\gamma< 3.8 \mbox{ GeV}$  
and $-t=0.45\pm0.05 \mbox{ GeV}^2$ (black), $-t=0.65\pm0.05 \mbox{ GeV}^2$ (red) and $-t=0.95\pm0.05 \mbox{ GeV}^2$ (blue). The error bars include both statistical and systematic uncertainties as explained in the text.}
\label{fig:final-4}
\end{figure}

\subsection{Method comparisons and final results}\label{sec:final_moments}
Moments derived by the different procedures agreed qualitatively.
The two  methods were consistent in the range of interest from 0.990 GeV$< M_{K^+K^-}< 1.075$~GeV   (and $0.6 < -t <  1.3 \mbox{ GeV}^2$). We do not use the region $M_{K^+K^-}>$ 1.075~GeV
to extract amplitude information because the choice of amplitude parametrization  (see Sec.~\ref{sec:amplitudes}) is only valid in proximity to  the $\phi(1020)$ meson mass. 
The difference between  the fit results of M1 and M2 was used to evaluate the systematic uncertainty 
associated with the moment extractions. 
The final results are  given as the  average of 
M1 (parametrization with moments) and M2 (parametrization with amplitudes), 
\begin{equation}
Y_{final}={{1}\over{2}}\sum_{i=1,2 \, Methods}{Y_i},
\end{equation}
where $Y$ stands for  $\langle Y_{\l\m} \rangle(E_\gamma,t,M_{K^+K^-})$.
The total uncertainty $\delta Y_{final}$ in the final moments was evaluated by adding in quadrature the statistical uncertainty, $\delta Y_{MINUIT}$ as given by MINUIT, and two systematic uncertainty contributions:
$\delta Y_{syst\,\, fit}$ related to the moment extraction procedure, 
and $\delta Y_{syst\,\,  norm}$, the systematic uncertainty  associated with the photon flux normalization (see Sec.~\ref{sec:exp}).
\begin{eqnarray}
\delta Y_{final}=\sqrt{\delta Y_{MINUIT}^2+\delta Y_{syst\,\, fit}^2+\delta Y_{syst\,\, norm}^2}\label{eq:err_Y}
\end{eqnarray}
with:\\
\begin{eqnarray}
\delta Y_{syst\,\,  fit}&=&\sqrt{\sum_{i=3,4 \, Methods}({Y_i}-Y_{final})^2}\\
\delta Y_{syst\,\,  norm}& =& 10\% \cdot Y_{final}.
\end{eqnarray}
Therefore, faor most of  the data points, the systematic uncertainties dominate over the statistical uncertainty. 
Samples of the  final experimental moments are shown in Figs.~\ref{fig:final-2},~\ref{fig:final-3}, and~\ref{fig:final-4}. The  error bars include the systematic uncertainties related to the moment extraction  and the photon flux normalization
as discussed in  Sec.~\ref{sec:final_moments}. The whole set of moments resulting from this analysis  is  available in the Jefferson Lab~\cite{jlab-db} and the Durham~\cite{dhuram-db} databases. 

\begin{figure}[h]
\begin{center}
\includegraphics[width = 3.2in]{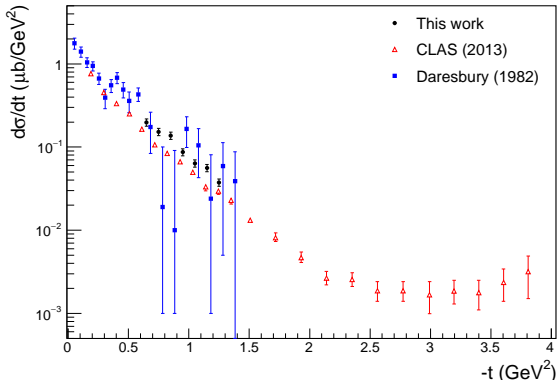} 
 \caption{Differential cross section for $ 3.0 <E_{\gamma} < 3.8 \mbox{ GeV}$ derived from the $\langle Y_{00} \rangle$ moment analysis compared with other results. 
 The differential cross section was calculated  from the $P$-wave extracted using the partial waves analysis described in  Sec.~\ref{sec:res_sec}. 
 The uncertainties include fit parameter uncertainties added in quadrature with a 10\% systematic uncertainty from the photon flux normalization. 
Results of this work are compared to CLAS published results from~ \cite{phi-clas-2014} 
in the energy range $ E_{\gamma} $= 3.300 GeV $\pm$ 0.015 GeV and Daresbury data~\cite{Barber} in the range $ 2.8 <E_{\gamma} < 3.8 \mbox{ GeV}$. }  
\label{fig:phi_xsec}
 \end{center}
 \end{figure}

As a check of the analysis procedure, the differential cross  section $d\sigma/dt$ for the $\gamma p \to p \phi(1020)$ meson  was extracted by integrating  the $\langle Y_{00} \rangle$ moment in each $t$ bin in the range  1.005 GeV  $<M_{K^+K^-}<$1.035 GeV  after subtracting a first-order polynomial background fitted to the data (excluding the region 1.005  GeV $< M_{K^+K^-} <$ 1.035 GeV as $\langle\tilde{Y}_{00}\rangle$ is not linear due to the $\phi$ peak).
The results are shown in Fig.~\ref{fig:phi_xsec}. 
Despite the different energy binning of the various studies, the reasonable agreement within the quoted uncertainties  with previous  measurements~\cite{Barber,phi-clas-2014}
gives us confidence in the accuracy of the analysis method.

\begin{figure*}
\includegraphics[width = 6.5in]{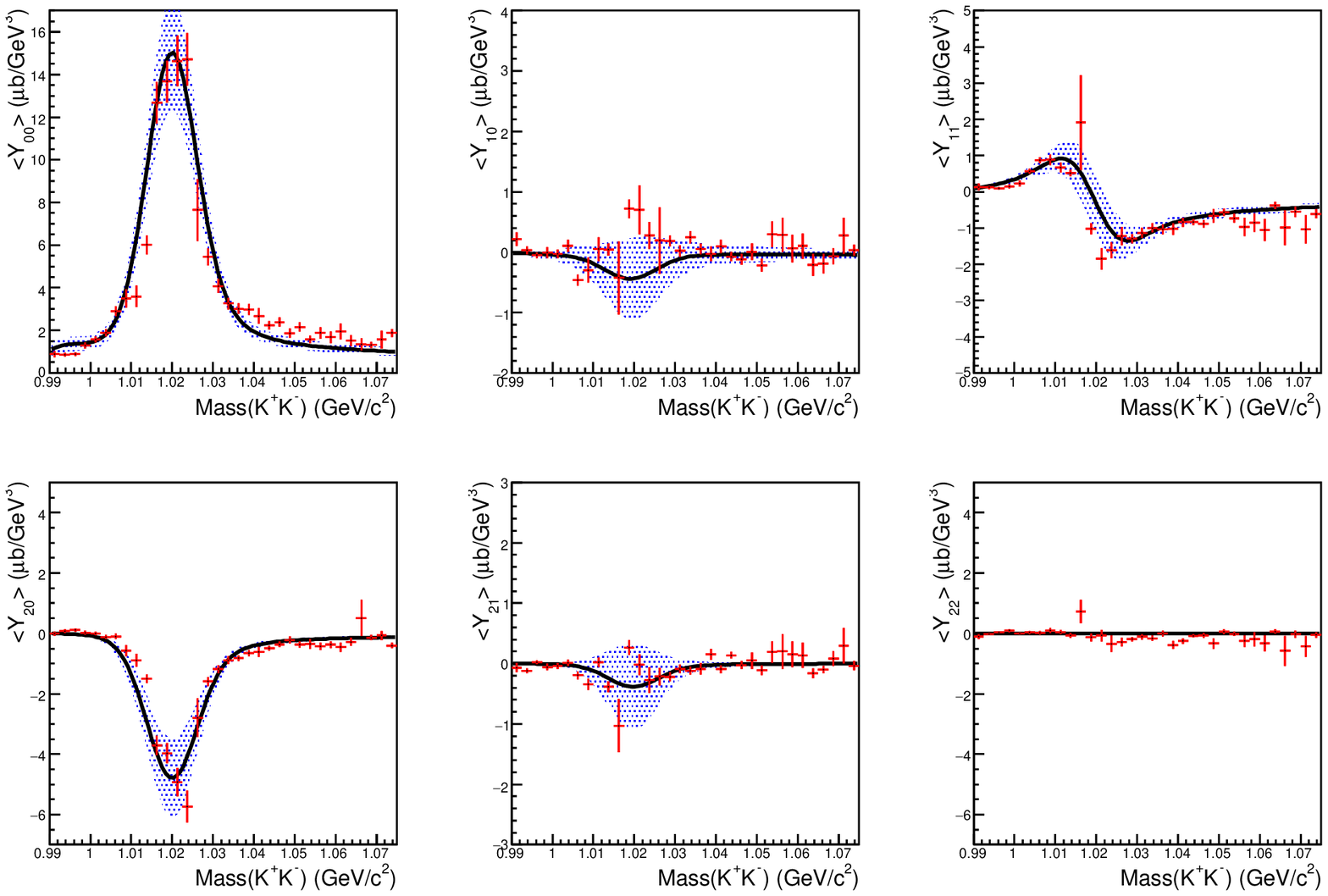}
\caption{Experimental moments $\langle Y_{LM} \rangle$ (red) for
            $0.6 \le \vert t \vert \le 0.7$~GeV$^2$ for $L \le 2$ and $M \le 2$ together
	    with the moments derived from the fitted amplitudes (black), including the
	    $L$=0 and $L$=1 amplitudes in the fit. The shaded band indicates the
	    associated systematic uncertainty. Under our assumptions (see text), $\langle Y_{22} \rangle = 0$ in the full mass range. The solid line represent the best fit.
} 
\label{fig:mom_results_start}
\end{figure*}

\section{\label{sec:disp} Partial wave analysis}
In the previous section we discussed how moments of the angular distributions of the $K^+K^-$ system, 
$\langle Y_{LM} \rangle$,  were extracted from the data in each bin in photon  energy, momentum transfer and di-kaon mass. 
In this section we describe how partial waves were parametrized and extracted by fitting the experimental moments.

 The production amplitudes can be written as
\begin{equation} 
f = f_{\lambda_\gamma,\lambda,\lambda'}(s,t,M_{K^+K^-},\Omega )= f_{\{\lambda\}}(s,t,M_{K^+K^-},\Omega).
\end{equation} 
where $\lambda_\gamma,\lambda,\lambda'$ are the helicities of the photon, target and recoil nucleon, respectively, and $M_{K^+K^-}$ is the invariant mass of the $K^+ K^-$ system. 
In terms of the helicity amplitudes the cross section is given by, 
\begin{equation}
 \frac{d\sigma}{d t d M_{K^+K^-} d \Omega} \;\left[ \frac{\mu b}{0.1\mbox{GeV}^2 2.5 \mbox{GeV}} \right]  = \Phi 
 |f_{\{\lambda\}}|^2
\end{equation}
with the phase space factor $\Phi$ given by 
\begin{equation} 
\Phi =  \frac{1}{4} \frac{1.5577}{64 \pi m_{N}^2E_{\gamma}^2} 
\frac{\sqrt{M_{K^+K^-}^2/4 - m_K^2}}
{2(2\pi)^3}, 
\end{equation} \\
 where the factor of $1/4$ comes from averaging over the initial photon and target polarizations and all dimensional quantities enter in 
 units of GeV.  The helicity amplitudes are decomposed into partial waves $f^{LM}_{\{\lambda\}}$ in the $K\bar K$ channel, 
\begin{equation} 
 f_{\{\lambda\}}(s,t,M_{K^+K^-},\Omega)  = \sum_{LM}  f^{LM}_{\{\lambda\}}(s,t,M_{K^+K^-}) Y_{LM}(\Omega).
 \label{pws}
 \end{equation}
 so that the  moments, defined in \eqref{eq:mom},
 are given by,

\begin{equation}
\frac{\langle Y_{LM}\rangle}{\Phi} = \sum_{L_1,M_1,L_2,M_2;\{\lambda\}} c_{L_1,M_1,L_2,M_2;LM} 
 \left[ f^{L_1M_1 *}_{\{\lambda\}} 
 f^{L_2M_2}_{\{\lambda\}}  \right].
\label{eq:explicit_moment_formula}
\end{equation}
with the $c$'s proportional to a product of Clebsch-Gordan 
 coefficients.    
Note that we are using the spherical basis for the spin projection $M$ and not the so-called reflectivity basis. 
 Equation \eqref{eq:explicit_moment_formula}  is a bilinear relation between the moments derived from the data and the partial wave amplitudes. The fit minimized the difference of the right  and the left side  of  Eq.~\eqref{eq:explicit_moment_formula}  with respect to free parameters in the amplitude parametrization.
 In this way, a set of moments was used to determine the amplitudes.

\begin{figure*}
\includegraphics[width = 6.5in]{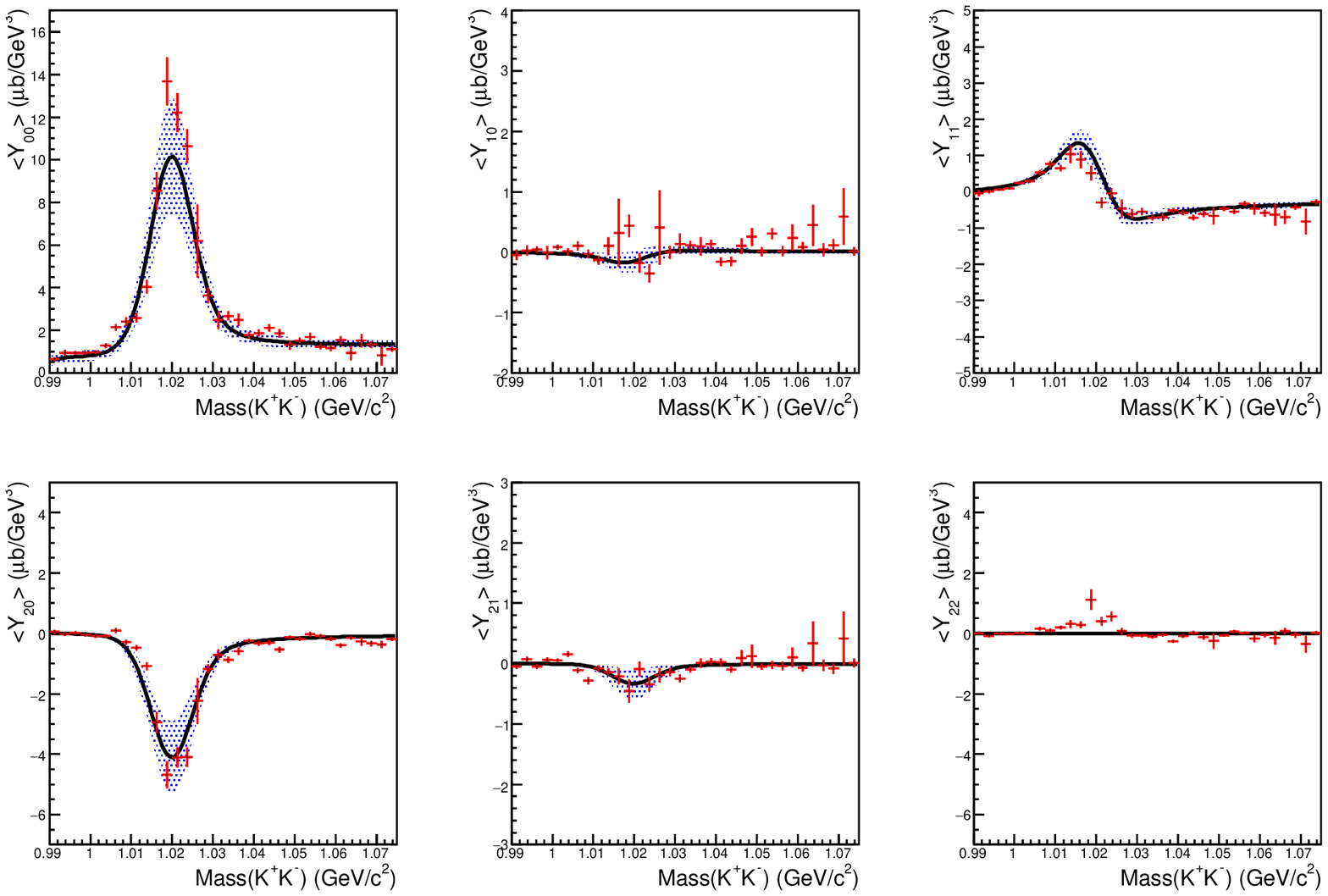}
\caption{Experimental moments $\langle Y_{LM} \rangle$ (red) for
            $0.7 \le \vert t \vert \le 0.8$~GeV$^2$ for $L \le 2$ and $M \le 2$ together
	    with the moments derived from the fitted amplitudes (black), including the
	    $L$=0 and $L$=1 amplitudes in the fit. The shaded band indicates the
	    associated systematic uncertainty. Under our assumptions (see text), $\langle Y_{22} \rangle = 0$ in the full mass range. The solid line represent the best fit.
}
\label{fig:results1}
\end{figure*}

\begin{figure*}
\includegraphics[width = 6.5in]{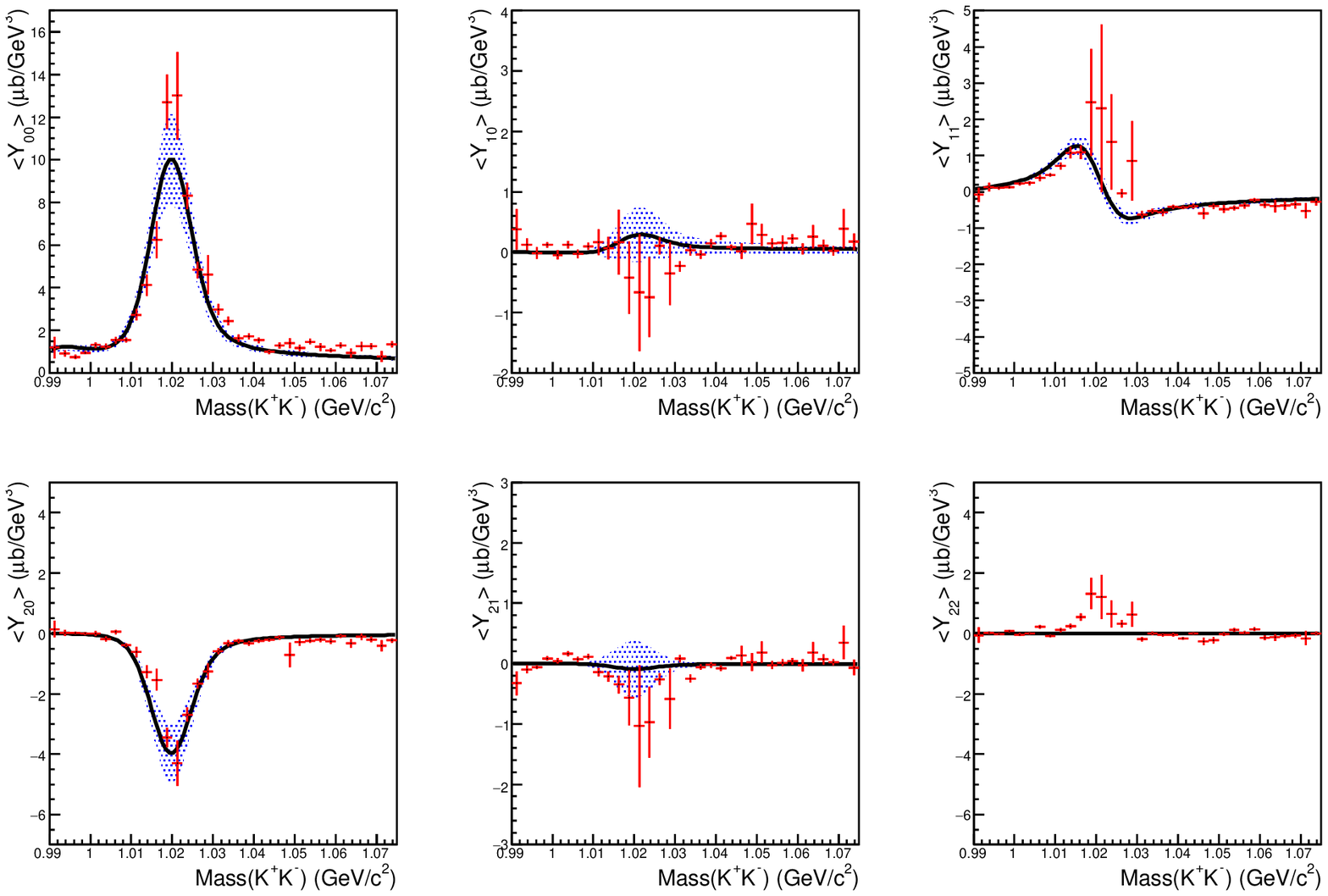}
\caption{Experimental moments $\langle Y_{LM} \rangle$ (red) for
            $0.8 \le \vert t \vert \le 0.9$~GeV$^2$ for $L \le 2$ and $M \le 2$ together
	    with the moments derived from the fitted amplitudes (black), including the
	    $L$=0 and $L$=1 amplitudes in the fit. The shaded band indicates the
	    associated systematic uncertainty. Under our assumptions (see text), $\langle Y_{22} \rangle = 0$ in the full mass range. The solid line represent the best fit.
}
\label{fig:results1}
\end{figure*}
\subsection{Parametrization of the partial waves}\label{sec:amplitudes}

 For a given $L$ and $M$, there are eight independent amplitudes, $f^{LM}_{\lambda_\gamma,\lambda,\lambda'} (M_{K^+K^-})$, in each energy and momentum transfer bin
corresponding to each combination of photon and initial and final nucleon helicity. 
We have only one energy bin in this analysis, so the fitted amplitudes do not depend on $E_\gamma$. Since the $L \ge 2$ amplitudes ($D$- and $F$-waves) are expected to be small in the $K^+K^-$ invariant mass range, we only include $S$- and $P$- partial-waves. The reaction $\gamma p \rightarrow p  K^+ K^- $
was then characterized by 32 amplitudes. There were 8 amplitudes required to describe the $S-$wave depending on the two spin projections of the photon ($\lambda_\gamma= \pm 1$), the target proton ($\lambda = \pm 1/2 $), and the recoil proton ($\lambda' = \pm 1/2$). In addition, there were 24 $P-$wave amplitudes depending also on 3 spin projections of the $\phi$. However, the photon helicity
 was restricted to  $\lambda_\gamma  = +1$
since the other amplitudes are related by parity conservation, resulting in 16 unconstrained amplitudes.
In addition, some  approximations in the parametrization  of the partial 
waves were adopted to reduce the number of free parameters in the fit as discussed below. In general, it is expected that the dominant amplitudes require minimal photon helicity flip, \textit{i.e.}
\begin{equation}
|f^{L1}| > |f^{L0}|.
 \end{equation}
 corresponding to photon helicity flip by zero and one,  respectively. In the $s$-channel helicity frame, we assume the $P$-wave production ($L=1$) is dominated by helicity non-flip amplitudes, {\it i.e.} the non-vanishing independent amplitudes are:
\begin{align}
P_{+} \equiv f^{1,1}_{+,+,+}, \ \ P_{-} \equiv  f^{1,1}_{+,-,-}, 
\end{align}
where $\pm$ refer to helicities of the photon and the protons, {\it e.g.} $+,+,+$ corresponds to $\lambda_\gamma = +1$, $\lambda = +1/2$ and $\lambda' = +1/2$.  
We introduced two additional amplitudes per each  orbital angular momentum, to describe unit photon helicity flip, 
\begin{align}
P_{0+} \equiv f^{1,0}_{+,+,+},\ \ P_{0-} \equiv f^{1,0}_{+,-,-} ,
\end{align}
and 
\begin{align} 
S_{+} \equiv f^{0,0}_{+,+,+},\ \
S_{-} \equiv f^{0,0}_{+,-,-} .
\end{align} 
  In the approximations described above, the dependence of moments on the $S$ and $P$ amplitudes  is given by, 

\begin{align} 
\langle Y_{00}\rangle &=  2[ |S_+|^2 + |S_-|^2 \nonumber \\
& + |P_+|^2 + |P_-|^2 + |P_{0+}|^2 + |P_{0-}|^2 ] \nonumber \\
\langle Y_{10} \rangle  &= 2   [S^*_+  P_{0+} + S^*_- P_{0-} + P^{*}_{0+} S_+ + P^{*}_{0-} S_-] \\ \nonumber
\langle Y_{11} \rangle  &=   P^*_+ S_+ + P^*_- S_-  + S^*_+ P_+ + S^*_- P_- \\ \nonumber
\langle Y_{20} \rangle  &= \frac{2}{\sqrt{5}} [ 2 |P_{0+}|^2 + 2 |P_{0-}|^2 -  |P_+|^2 -  |P_-|^2 ] \\ \nonumber
\langle Y_{21} \rangle  &= \sqrt{\frac{3}{5}} [  P^*_{0+} P_+ + P^*_{0-} P_-  + P^*_+ P_{0+} + P^*_- P_{0-} ].
\label{explicit_moment_formula}
\end{align} 
with $\langle Y_{22} \rangle$ vanishing under our assumptions.
 Here we see the $\langle Y_{10} \rangle$ and $\langle Y_{11} \rangle$ moments contain information about the presence of the  $S$-wave interference with the dominant $P$-wave. Thus a nonzero  $\langle Y_{10} \rangle$ or $\langle Y_{11} \rangle$ moment is an indication of a non-vanishing  $S$-wave amplitude. 
   In order for the $\langle Y_{22} \rangle$ moment to be non-zero, there must be two-unit photon helicity flip amplitudes. Given that there is no significant structure in any $\langle Y_{22} \rangle$ moments of this analysis, it is justified to neglect two-unit photon helicity flip amplitudes.
 So far we have introduced only the nucleon helicity non-flip amplitudes.  Indeed $P$-wave nucleon helicity flip amplitudes are expected to be small ({\it cf. } Appendix~\ref{appendix:amp} and Ref.~\cite{Lesniak:2003}).  

Without polarization information, it is difficult to separate out amplitudes differing only by the helicity of the nucleon.
We did attempt to fit the data using various configurations of nucleon helicity amplitudes and found in particular that the $S-P$ interference signal in the $\langle Y_{11} \rangle$ moment cannot be described solely by interference between nucleon flip amplitudes.
 We comment on this further in Sec.~\ref{pwa}.  We find, however,  that the moments can be well described by interference between the dominant, nucleon helicity non-flip $P$- and $S$-wave amplitudes.  
Details of the amplitude  parametrization are given in Appendix~\ref{appendix:amp}.

\subsection{Fit of the moments}

\begin{figure*}
\includegraphics[width = 6.5in]{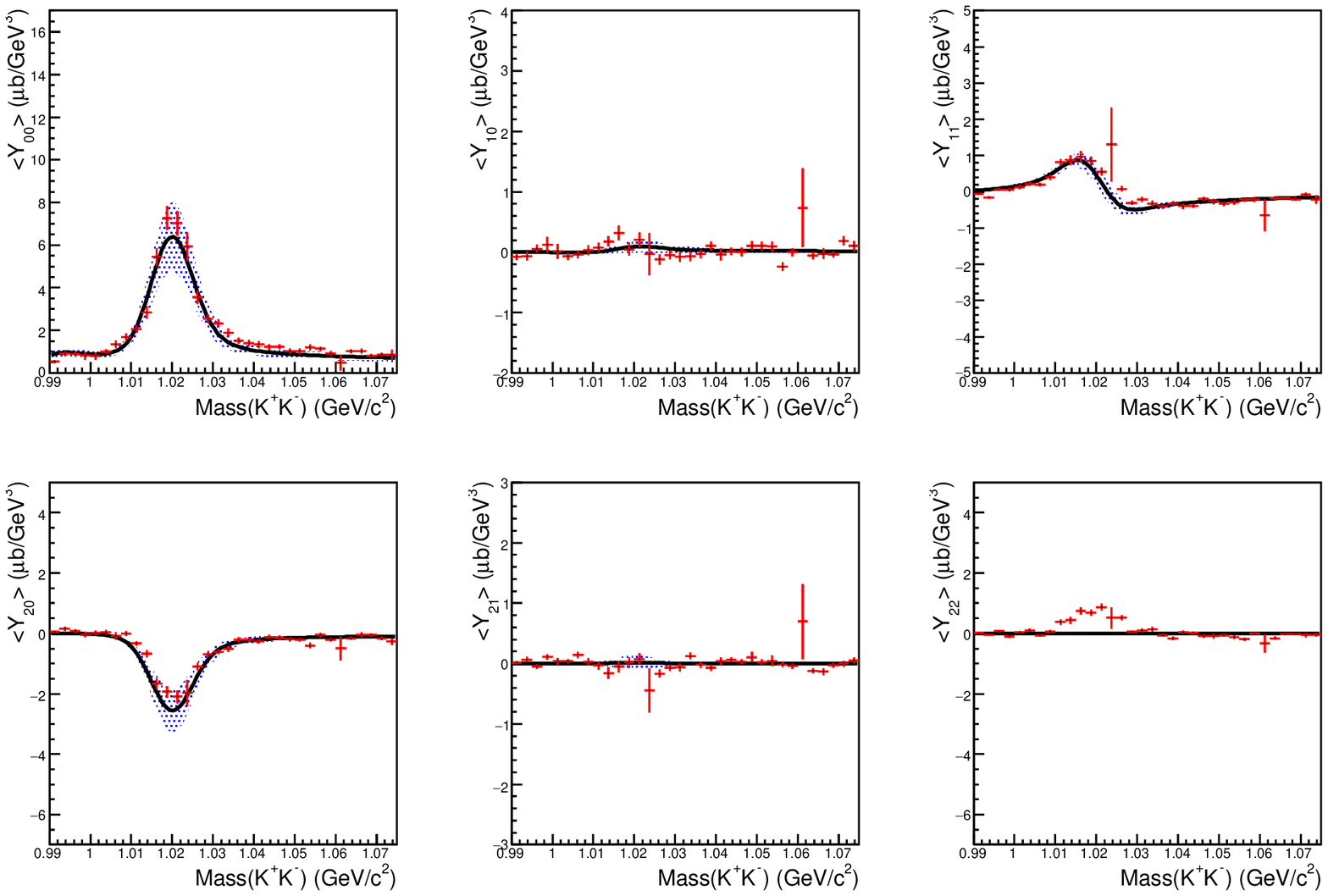}
\caption{Experimental moments $\langle Y_{LM} \rangle$ (red) for
            $0.9 \le \vert t \vert \le 1.0$~GeV$^2$ for $L \le 2$ and $M \le 2$ together
	    with the moments derived from the fitted amplitudes (black), including the
	    $L$=0 and $L$=1 amplitudes in the fit. The shaded band indicates the
	    associated systematic uncertainty. Under our assumptions (see text), $\langle Y_{22} \rangle = 0$ in the full mass range. The solid line represent the best fit.
}
\label{fig:results1}
\end{figure*}
\begin{figure*}
\includegraphics[width = 6.5in]{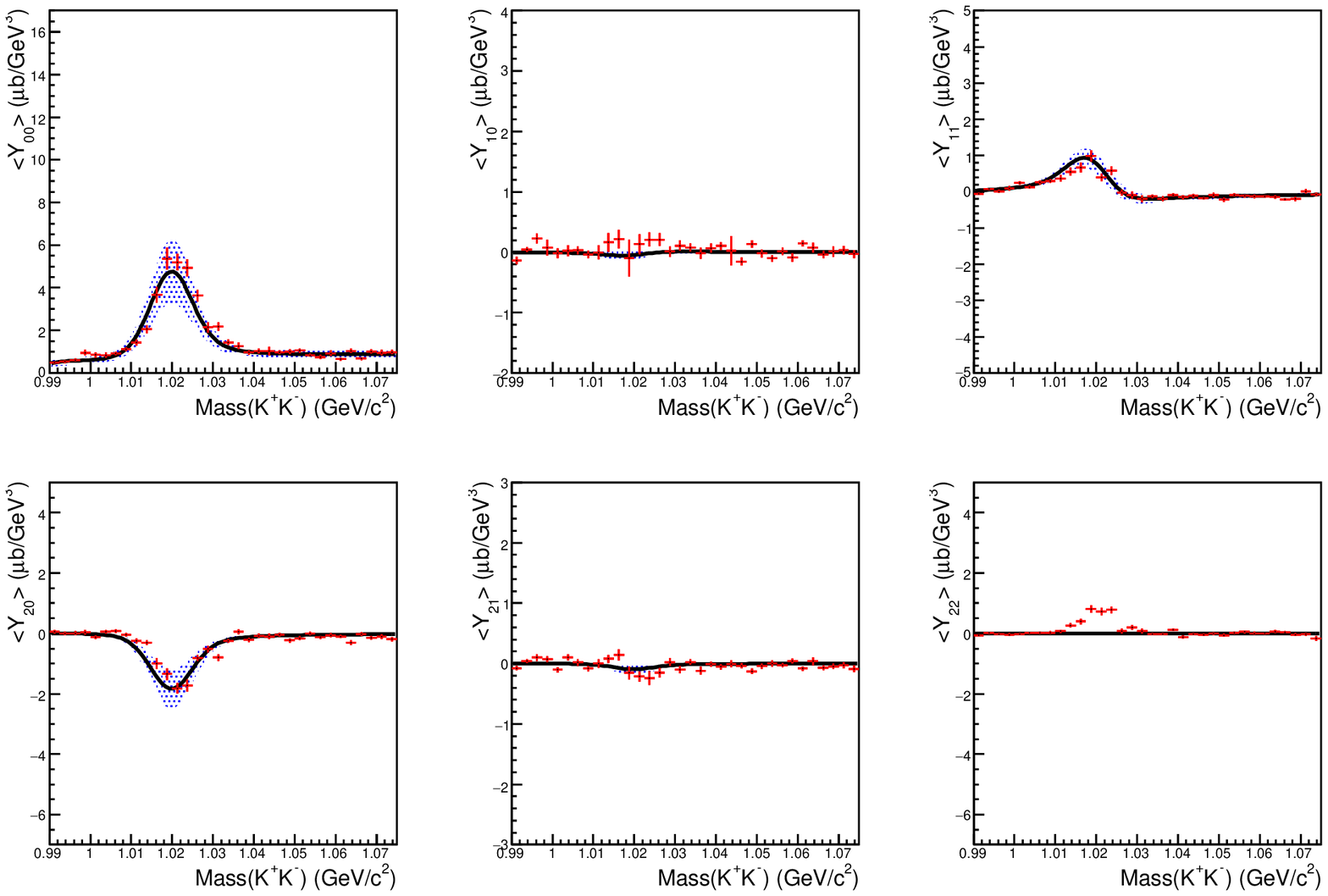}
\caption{Experimental moments $\langle Y_{LM} \rangle$ (red) for
            $1.0 \le \vert t \vert \le 1.1$~GeV$^2$ for $L \le 2$ and $M \le 2$ together
	    with the moments derived from the fitted amplitudes (black), including the
	    $L$=0 and $L$=1 amplitudes in the fit. The shaded band indicates the
	    associated systematic uncertainty. Under our assumptions (see text), $\langle Y_{22} \rangle = 0$ in the full mass range. The solid line represent the best fit.
}
\label{fig:results1}
\end{figure*}
\begin{figure*}
\includegraphics[width = 6.5in]{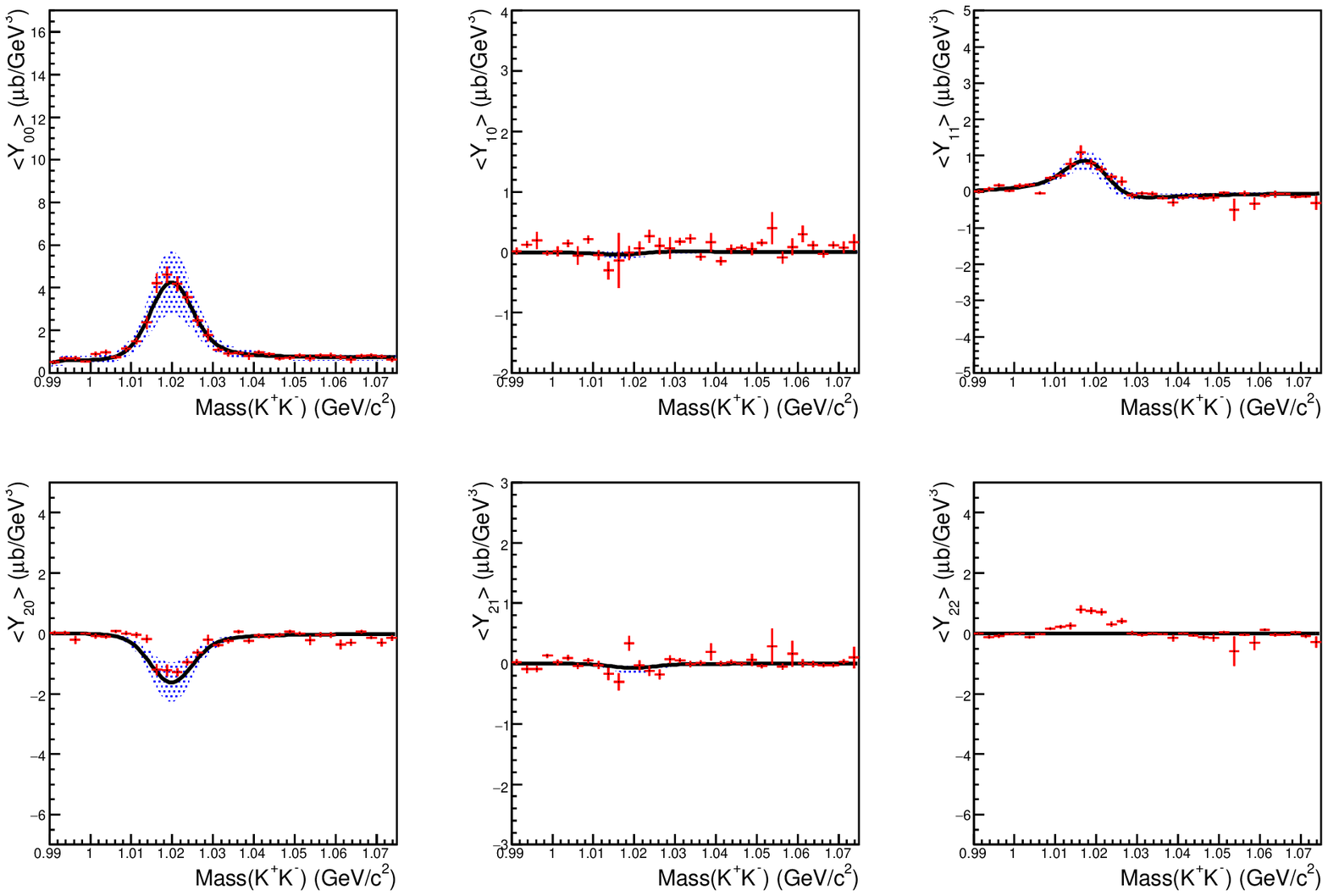}
\caption{Experimental moments $\langle Y_{LM} \rangle$ (red) for
            $1.1 \le \vert t \vert \le 1.2$~GeV$^2$ for $L \le 2$ and $M \le 2$ together
	    with the moments derived from the fitted amplitudes (black), including the
	    $L$=0 and $L$=1 amplitudes in the fit. The shaded band indicates the
	    associated systematic uncertainty. Under our assumptions (see text), $\langle Y_{22} \rangle = 0$ in the full mass range. The solid line represent the best fit.
}
\label{fig:results1}
\end{figure*}
\begin{figure*}
\includegraphics[width = 6.5in]{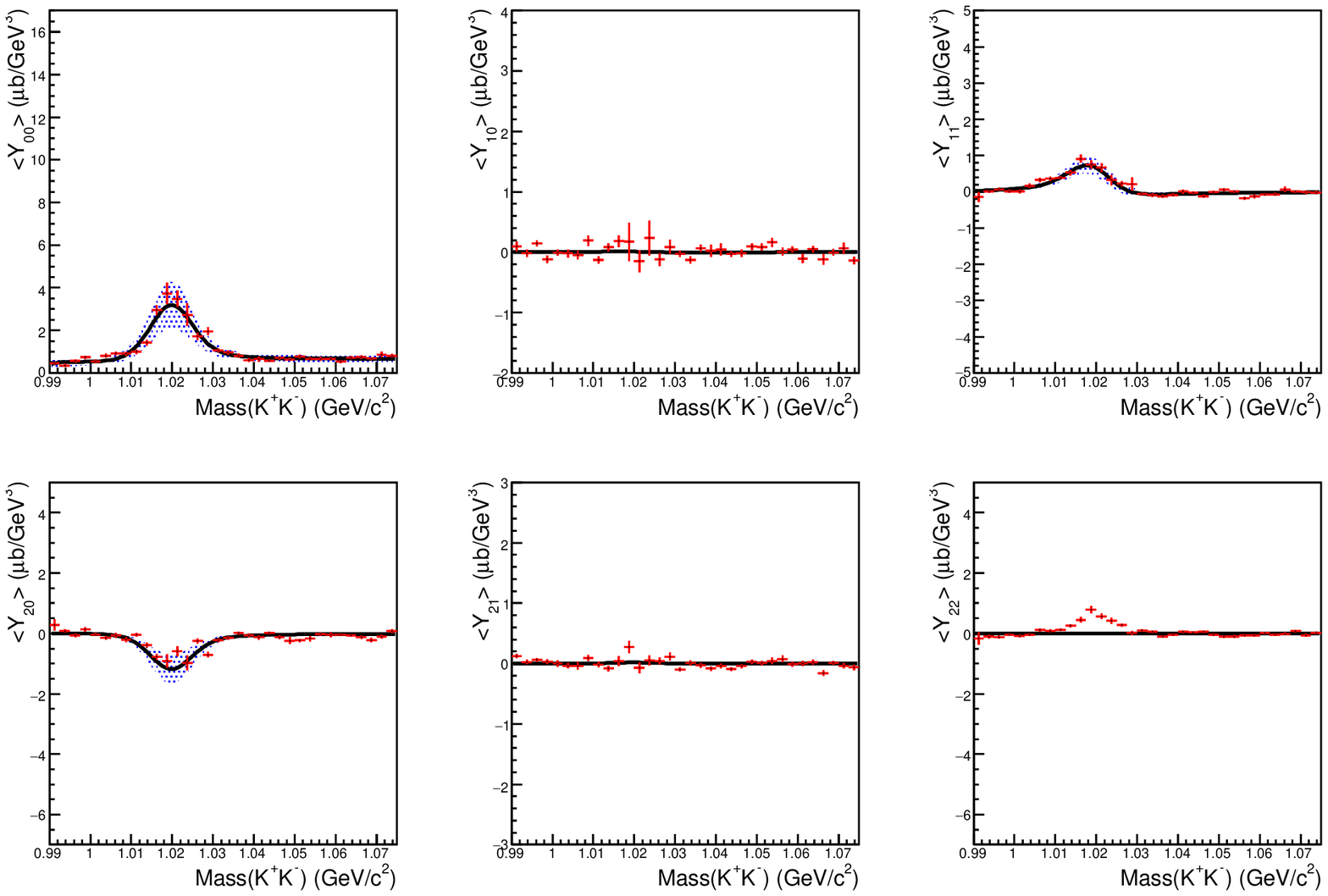}
\caption{Experimental moments $\langle Y_{LM} \rangle$ (red) for
            $1.2 \le \vert t \vert \le 1.3$~GeV$^2$ for $L \le 2$ and $M \le 2$ together
	    with the moments derived from the fitted amplitudes (black), including the
	    $L$=0 and $L$=1 amplitudes in the fit. The shaded band indicates the
	    associated systematic uncertainty. Under our assumptions (see text), $\langle Y_{22} \rangle = 0$ in the full mass range. The solid line represent the best fit. }
\label{fig:mom_results_end}
\end{figure*}

To account for detector resolution, the moments calculated from the amplitudes were smeared by a Gaussian function. The $\phi$ width apparent in the $\langle Y_{00} \rangle$ moment determined the smearing needed in order for the $P$-wave parametrization (with fixed $\phi$ width) to match the data.  This lead to a width in the Gaussian 
smearing of  4 MeV,  which is compatible with the CLAS detector resolution measured in other reactions~\cite{devita}. 
We fit the moments $\langle Y_{LM} \rangle$ with $L \le 2$ and $M \le 2$ using up to $L=1\ (P)$ waves as described above. 
In Figs. \ref{fig:mom_results_start} - \ref{fig:mom_results_end}, we present the fit results of this analysis from $0.6 < -t < 1.3$  GeV$^2$. To properly take into account the uncertainty contributions (statistical and systematic) to the experimental moments described in Sec.~\ref{sec:final_moments}, the two sets of moments from methods M1 and M2 were individually fit, and the fit results were averaged, obtaining the central value shown by the black line in the figures. The error band, shown as a grey area, was calculated following the same procedure adopted for the experimental moments (Sec.~\ref{sec:final_moments}). The two lowest momentum transfer bins $ 0.4 \le t \le 0.6  \mbox{ GeV}^2$ were excluded from the analysis because the moment reconstruction procedure was found not to be reliable in this region. In addition, the $\langle Y_{10} \rangle$ moment was not used to extract the $S$-wave magnitude because the  procedure could not  always reproduce an accurate $\langle Y_{10} \rangle$ moment based on  tests  performed on pseudo-data.

\subsection{Partial wave amplitudes} \label{pwa} 
\begin{figure}
\includegraphics[width = 3.5in]{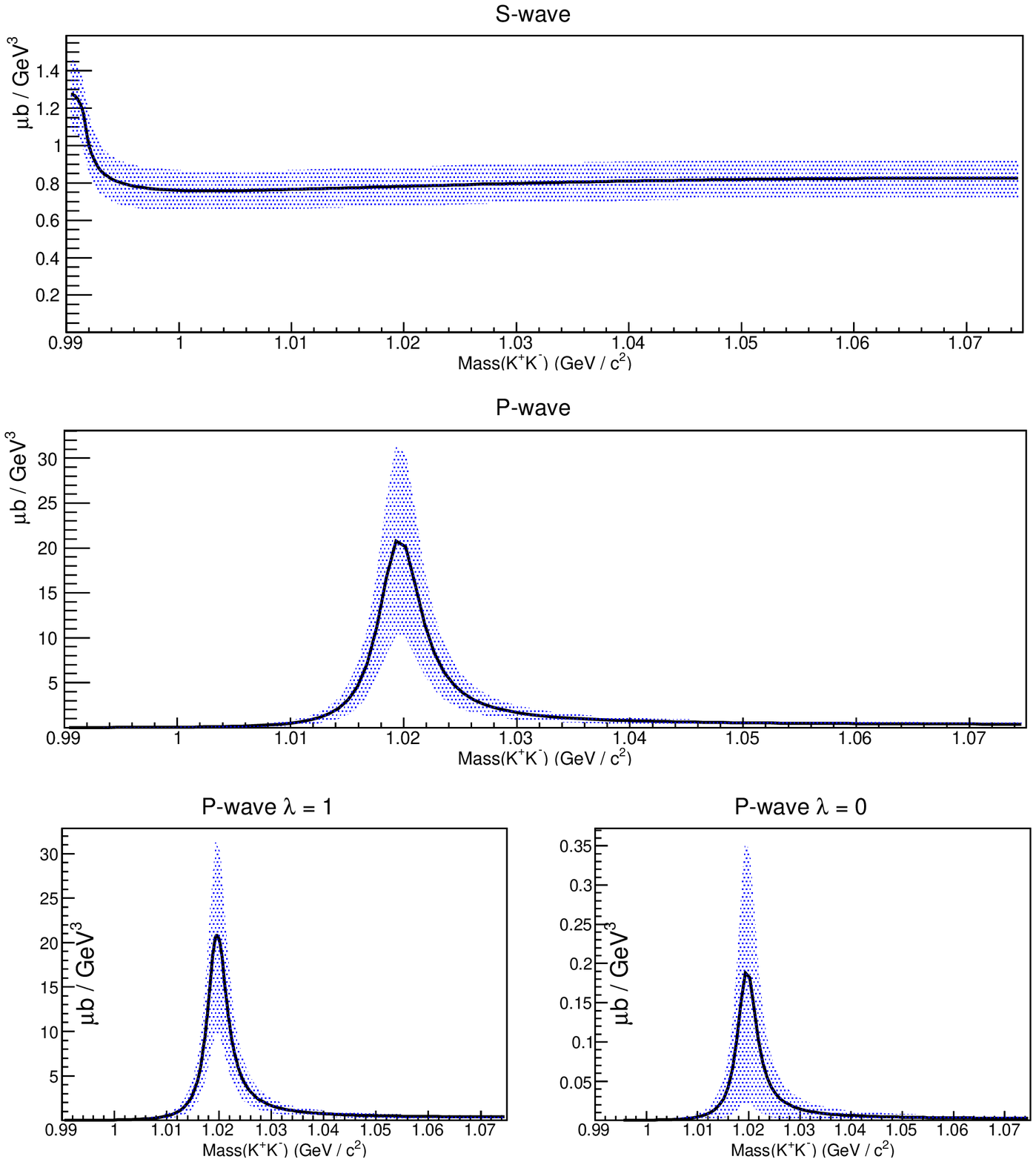}
\caption{Magnitudes of the $S$ and $P$ partial waves along with two spin projections of the $P-$wave ($\lambda_{KK}=1,0$) in the $0.7 \le t \le 0.8 $ GeV$^2$ bin determined by fitting to the experimental moments. }
\label{fig:waves_res}
\end{figure}

\begin{figure}
\includegraphics[width = 8.05cm]{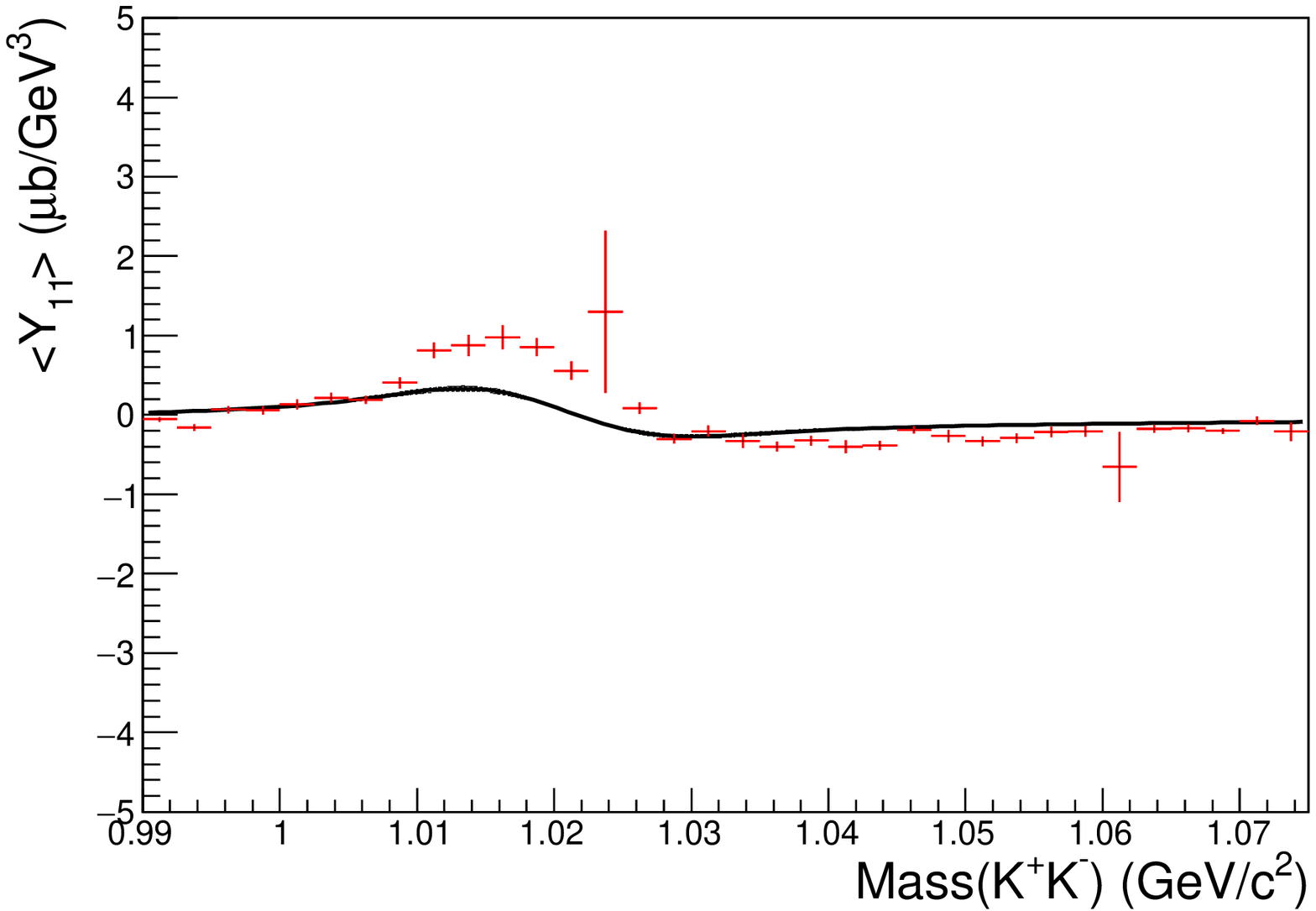}
\caption{Fit of $\langle Y_{11} \rangle$ moment with the nucleon flip amplitude alone. The bad agreement indicates  the non-flip amplitudes dominate  the measured moments.}
\label{fig:flip_test}
\end{figure}

As an example, the square of the magnitude of the $S$- and $P$ partial-waves derived by fit for the momentum transfer bin 
$0.7 < -t < 0.8$ GeV$^2 $ are shown in Fig. \ref{fig:waves_res}. The $S$-wave threshold enhancement provides a hint of the scalar $f_0(980)$ or $a_0(980)$ states, which have been parametrized by the exchange
of the $\omega$ and $\rho$ vector mesons in the $t$-channel. The top and the middle plots show the partial waves summed over all helicities. The two bottom plots show the amplitudes for two possible values of $M=1,0$, the helicity of the di-kaon system. Note that we use the wave with photon helicity  $\lambda_\gamma=+1$ as a reference. Thus, $M = 1$ corresponds to the no-helicity flip ($s-$channel helicity conserving) amplitude, which, as expected, is the dominant one, and $M=0$ corresponds to unit photon helicity flip. 
The non-vanishing $\langle Y_{22} \rangle$ moments show the presence of a small two unit helicity flip amplitude. By neglecting  the $M=-1$ amplitudes, we have focused on describing the dominant structure  in the $\langle Y_{11} \rangle$ and $\langle Y_{20} \rangle$ moments and reducing the number of fit parameters. \\
To check sensitivity to various helicity components we performed  the fit in three configurations. In the first configuration we included  $S$- and $P$-wave amplitudes with vanishing photon helicity flip and unit photon helicity flip. Nucleon helicity flip amplitudes were excluded.   In the second configuration, we used Regge factorization to reduce the number of independent amplitudes. Specifically,  the parity relation applied to the nucleon vertex~\cite{Irving:1977ea} reduces the number of unconstrained amplitudes by a factor of two, since $S_+$ is related to $S_-$, $P_+$ to $P_-$, and $P_{0+}$ to $P_{0-}$. Finally in the third configuration we used the above Regge-constrained $P$-wave amplitudes and we added to them the nucleon helicity flip amplitudes. In this configuration we tested if the interference signal in the moments could be described by interfering nucleon flip amplitudes by attempting to extract the nucleon helicity flip amplitudes from the $\langle Y_{10} \rangle$ and $\langle Y_{11} \rangle$ moments. 
 Specifically, we added two nucleon helicity flip $P-$wave amplitudes
$ f^{1,1}_{++-}$, $f^{1,0}_{++-}$  and one nucleon flip $S-$wave amplitude $f^{0,0}_{++-}$. It is only necessary to consider one-half of all the nucleon flip amplitudes because the others are not independent after using the Regge factorization condition. 
We found  that the first two configurations gave similar results, and specifically, in Figs.~\ref{fig:mom_results_start}-\ref{fig:mom_results_end}, we show the results obtained with the 
 second configuration described above.
In the third  configuration 
a fit was first performed using the  $\langle Y_{00} \rangle$ and $\langle Y_{20} \rangle$  moments to  extract the dominant nucleon non-flip $P-$wave, while setting the nucleon flip amplitudes to zero.
After fixing the strength of the non-flip $P$-wave in this way, we introduced nucleon flip $P-$ and $S$-waves and added the $\langle Y_{10} \rangle$ and $\langle Y_{11} \rangle$  moments to the fit. As shown in Fig.~\ref{fig:flip_test} , we found the nucleon flip amplitudes cannot be large enough to significantly affect the $\langle Y_{11} \rangle$  moment. We thus conclude that the non-flip amplitudes dominate the measured moments. 

\subsection{Differential cross sections} \label{sec:res_sec}
\begin{figure}
\includegraphics[width = 8cm]{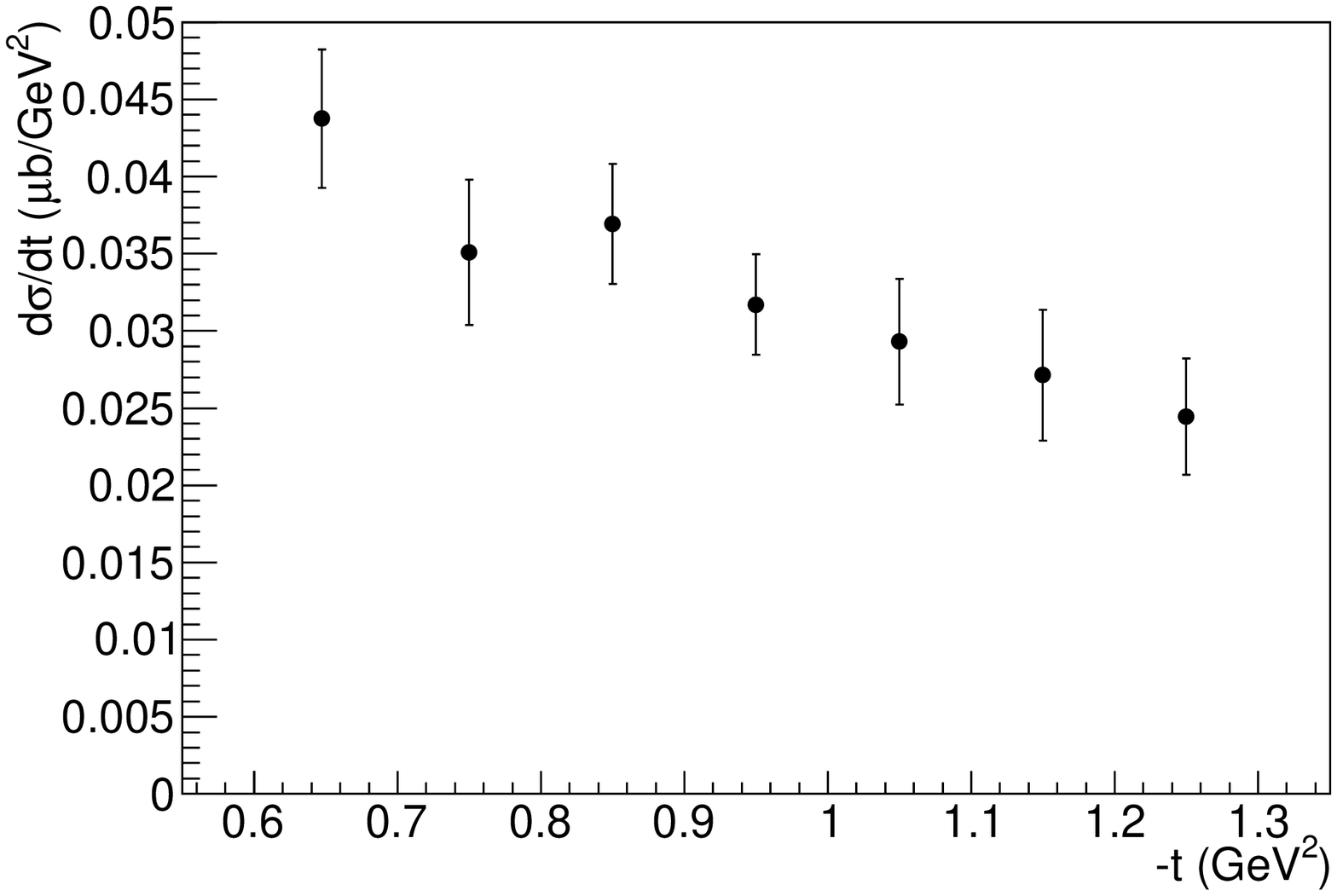}
\caption{Differential cross section obtained from integrating the $S-$wave magnitude in the $M_{K^+K^-}$ range $1.0195 \pm 0.0225$ GeV }
\label{fig:dsigma1}
\end{figure}

\begin{figure}
\includegraphics[width = 8cm]{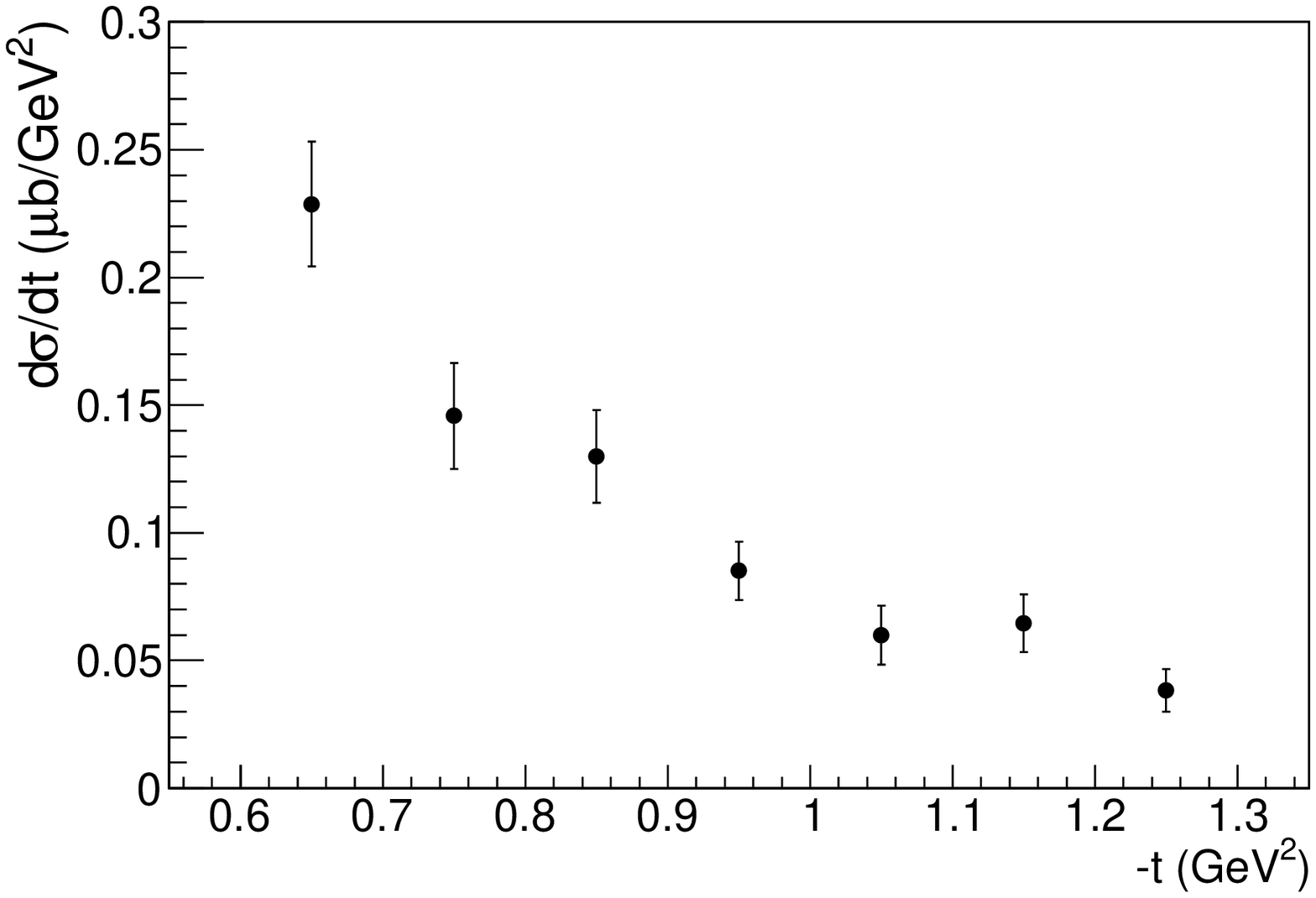}
\caption{Differential cross section obtained from integrating the $P-$wave magnitude in the $M_{K^+K^-}$ range $1.0195 \pm 0.0225$ GeV.  }
\label{fig:dsigma2}
\end{figure}

Differential cross section $(d\sigma/dt)_L$ for individual waves
 can be obtained by integrating the corresponding amplitude obtained from fits to the moments. The rCrossesults  are shown in Figs. \ref{fig:dsigma1} and \ref{fig:dsigma2}.  All cross sections are found by integrating the mass region $1.0195 \pm 0.0225$ GeV.  
It is worth noting that the  magnitudes of the $S$ and $P_0$ waves found in this analysis (see Table \ref{tab:results}) are consistent with predictions (summarized in Table  \ref{tab:lesniak}) of a model constrained on a somewhat higher photon energy data~\cite{Barber,Fries78,Behrend}. The discrepancy can be explained by the different $-t$ integration range.

 \begin{table}
\centering
\begin{tabular}{|c|c|}
	\hline
	photon energy & 3.0 - 3.8 $\mbox{ GeV}$ \\
	\hline
total cross section & 27.2 \\
	\hline
sum of $P$-waves & 22.9 $\pm$ 2.4    \\
	\hline
$P_0$-wave   & 1.9 $\pm$ 0.6   \\
	\hline
$S$-wave &    4.3 $\pm$ 0.45 \\
	\hline
\end{tabular}
\caption{ Cross sections in nb
 obtained from this analysis by integrating the $S$- and $P$-wave  magnitudes in the $M_{K^+K^-}$ range $1.0195 \pm 0.0225$ GeV in the single momentum transfer bin $0.6 \le - t \le 0.7 \mbox{ GeV}^2$.}
\label{tab:results}
\end{table}

 \begin{table}
\centering
\begin{tabular}{|c|c|c|}
	\hline
	photon energy & 4.00 $\mbox{ GeV}$ & $5.65 \mbox{ GeV}$\\
	\hline
sum of $P$-waves & $218.4 \pm 39.5$ & $120.5 \pm 9.4$     \\
	\hline
 background & $300.0^{+10.0}_{-10.7}$ & $4.7^{+4.2}_{-5.8}$ \\
  \hline 	
$P_0$-wave   & $4.7^{+5.7}_{-4.5}$ &  $14.0^{+5.3}_{-4.8}$ \\
	\hline
$S$-wave &    $4.3^{+6.6}_{-3.6}$ & $6.8^{+6.6}_{-4.3}$ \\
	\hline
\end{tabular}
\caption{ Cross sections in nb  obtained from integrating the $S-$  and $P-$waves from the Regge model of  \cite{Lesniak:2005}. The results shown are 
 integrated over $-t$ up to  $1.5\mbox{ GeV}^2$ and the  
$M_{K\bar{K}}$ range of $(0.997-1.042)$ GeV for 
$E_\gamma = 4\mbox{ GeV}$ and up to $-t$ of 
$0.2\mbox{ GeV}^2$ an $M_{K\bar K}$ in the range $(1.01-1.03)$ GeV at $E_\gamma = 5.65\mbox{ GeV}$, respectively. 
}
\label{tab:lesniak} 
\end{table}


\subsection{Uncertainty  evaluation}
The final uncertainty was computed as the sum in quadrature of the statistical uncertainty of the fit, and two systematic uncertainty  contributions: the first related to the moment extraction procedure, evaluated as the variance of the two fit results, and the second associated with the photon flux normalization estimated to be 10\%. The central values and uncertainties for all of the observables of interest discussed in the next sections were derived from the fit results with the same procedure.

\section{\label{sec:sum} Summary}
In summary, we performed a partial wave analysis of the reaction  $\gamma p \to p K^+ K^-$ in the photon energy range 3.0-3.8~GeV
and momentum transfer range $-t=0.6-1.3$~GeV$^2$. 
Peripheral photoproduction of meson resonances is an important reaction to study their structure.
 On one side, photons have a  point-like coupling to quarks, which enhances production of compact states. On the other, pion exchange amplitudes in photoproduction on the nucleon can be used to determine rate of resonance production through final state interactions.  Theoretical analysis of these process are currently underway \cite{Bibrzycki:2013pja}.
Moments of the di-kaon angular distributions, defined as bi-linear functions of the partial wave amplitudes,
were fitted to the experimental data by means of an un-binned likelihood procedure. Different parametrisation bases were used and detailed systematic checks 
were performed to ensure the reliability of the analysis procedure. We extracted moments $\langle Y_{LM}\rangle$ with $L \le 4$ and $M \le 2$ by using amplitudes with $L \le 2$ (up to $P$-waves). 
The production amplitudes have been parametrized using a Regge-theory inspired model. The $P-$wave, dominated by the $\phi(1020)$-meson, was parametrized by  Pomeron exchange, while the $f_0(980)$ meson  in the  $S$-wave was described by the exchange of the $\omega$ and $\rho$ vector mesons in the $t$-channel. This model also accounts for the final state interaction (FSI) of  the emitted kaons. 
The moment $\langle Y_{00} \rangle$ is dominated by the $\phi(1020)$ meson contribution in the $P$-wave,
while the moments $\langle Y_{10} \rangle$ and $\langle Y_{11} \rangle$  show  contributions of the $S$-wave  through 
interference with the $P$-wave. The cross sections of $S$- and $P$-waves in the mass range of the $\phi(1020)$,  were 
computed. This is the first time the  $t$-dependent cross section of the $S$-wave contribution to the elastic $K^+K^-$ photoproduction has been measured.

\section{\label{sec:ack} Acknowledgments}
We would like to acknowledge the outstanding efforts of the staff of the Accelerator
and the Physics Divisions at Jefferson Lab that made this experiment possible. 
 This work was supported in part by 
 the Chilean Comisi\'on Nacional de Investigaci\'on Cient\'ifica y Tecnol\'ogica (CONICYT),
 the Italian Istituto Nazionale di Fisica Nucleare,
 the French Centre National de la Recherche Scientifique,
 the French Commissariat \`{a} l'Energie Atomique,
 the U.S. Department of Energy,
 the National Science Foundation,
 the Scottish Universities Physics Alliance (SUPA),
 the United Kingdom's Science and Technology Facilities Council,
 and the National Research Foundation of Korea.
 The Southeastern Universities Research Association (SURA) operates the
 Thomas Jefferson National Accelerator Facility for the United States
 Department of Energy under contract DE-AC05-84ER40150.
This material is based
upon work supported by the U.S. Department of Energy, Office of Science,
Office of Nuclear Physics under contract DE-AC05-06OR23177.

\appendix 
\section{Systematic studies of the  moment extraction}\label{appendix:syst}

\subsection{Energy bin size}
\begin{figure*}
\includegraphics[width=6in]{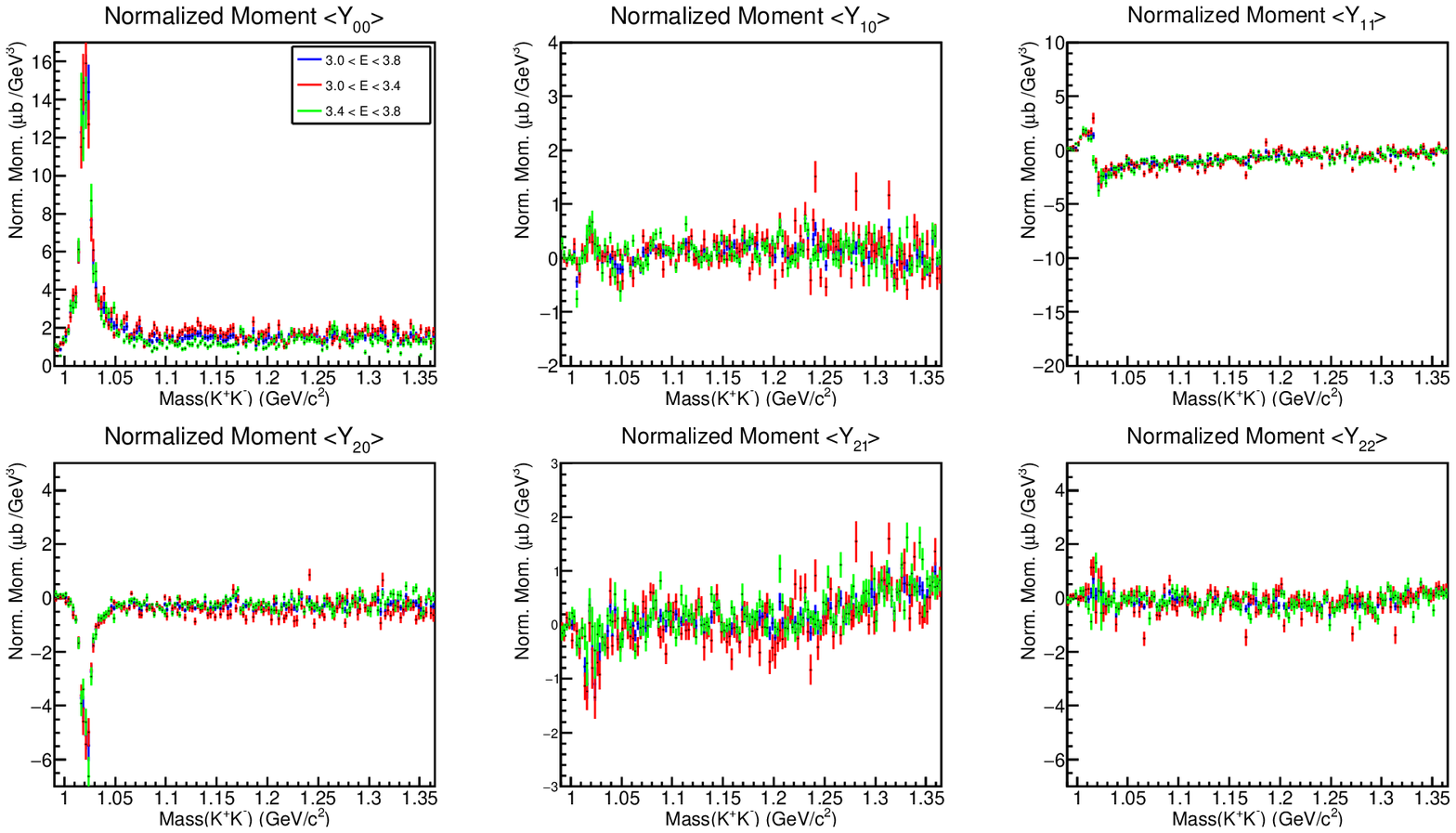}
\caption{Normalized moments obtained from method M1 with varying energy bin sizes and $\lambda_{max}=2$. Blue corresponds to the full energy range  3.0 GeV $ < E_\gamma < $3.8 GeV. Red and green corresponds to 
3.0 GeV $<  E_\gamma <  $3.4 GeV and 
3.4 GeV $ < E_\gamma < $ 3.8 GeV, respectively.}
\label{fig:ebin1}
\end{figure*}

Two energy bin configurations were studied: a single bin with $3.0 < E_{\gamma} < 3.8 \mbox{ GeV}$ and two bins $3.0 < E_{\gamma} < 3.4 \mbox{ GeV} $ and $3.4 < E_{\gamma} < 3.8 \mbox{ GeV}$. The moments were  more stable for the single energy bin configuration due to larger statistics. However, the kaon-nucleon mass distributions were better reproduced using the smaller bin size. The angular moments obtained from both configurations are shown to be in good agreement in Fig. \ref{fig:ebin1}.

\subsection{Cut on $M_{K^-p} > 1.6$  GeV }
\begin{figure}[htpb]
\includegraphics[width=3.5in]{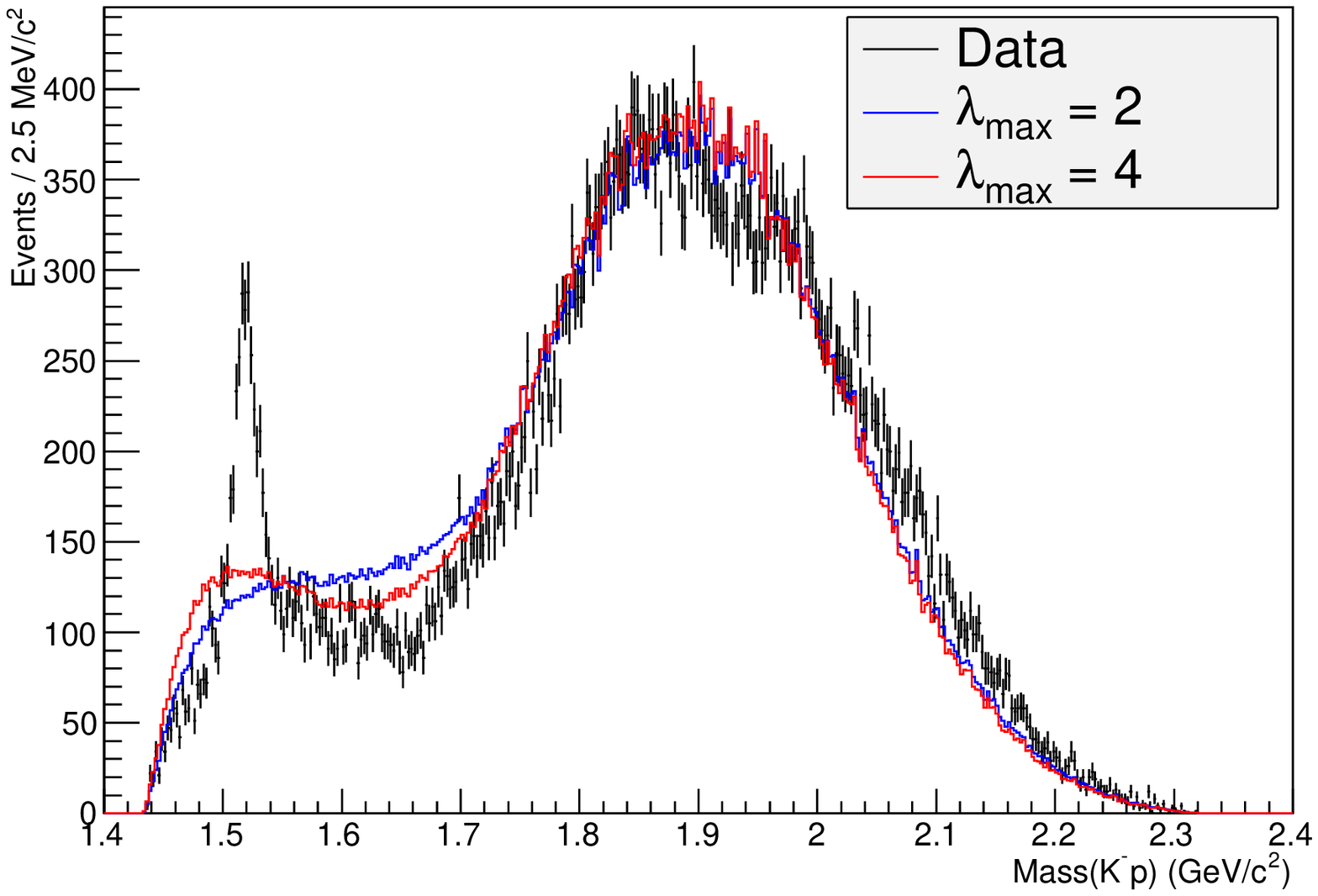}
\caption{Measured number of events as a function of the $pK^-$ invariant mass compared to the predicted distribution computed with fitted results from method 3 weighted by the experimental acceptance before cutting out the $\Lambda(1520)$.}
\label{fig:before_cut1}
\end{figure}

The $\Lambda(1520)$ peak in the $K^-p$ mass distribution cannot be reproduced with $\lambda_{max} < 4$ with any of the four methods. Fig. \ref{fig:before_cut1} shows the  fit results before cutting out the region containing the $\Lambda(1520)$.
This region is not a main focus of this study, so the kinematical region with $M_{K^-p} < 1.6$  GeV was removed from this analysis. Dalitz plots of the whole  $p K^+ K^-$ data set before and after this cut, show that the number of events in the $M_{K^+K^-}$ region near the $\phi$ mass were not affected by this cut. Therefore, the systematic effect of this cut on the determined cross sections is negligible. 

\subsection{Sensitivity to $\lambda_{max}$ and effect of truncation to $\lambda_{max}=4$}
\begin{figure*}
\includegraphics[width=6.5in]{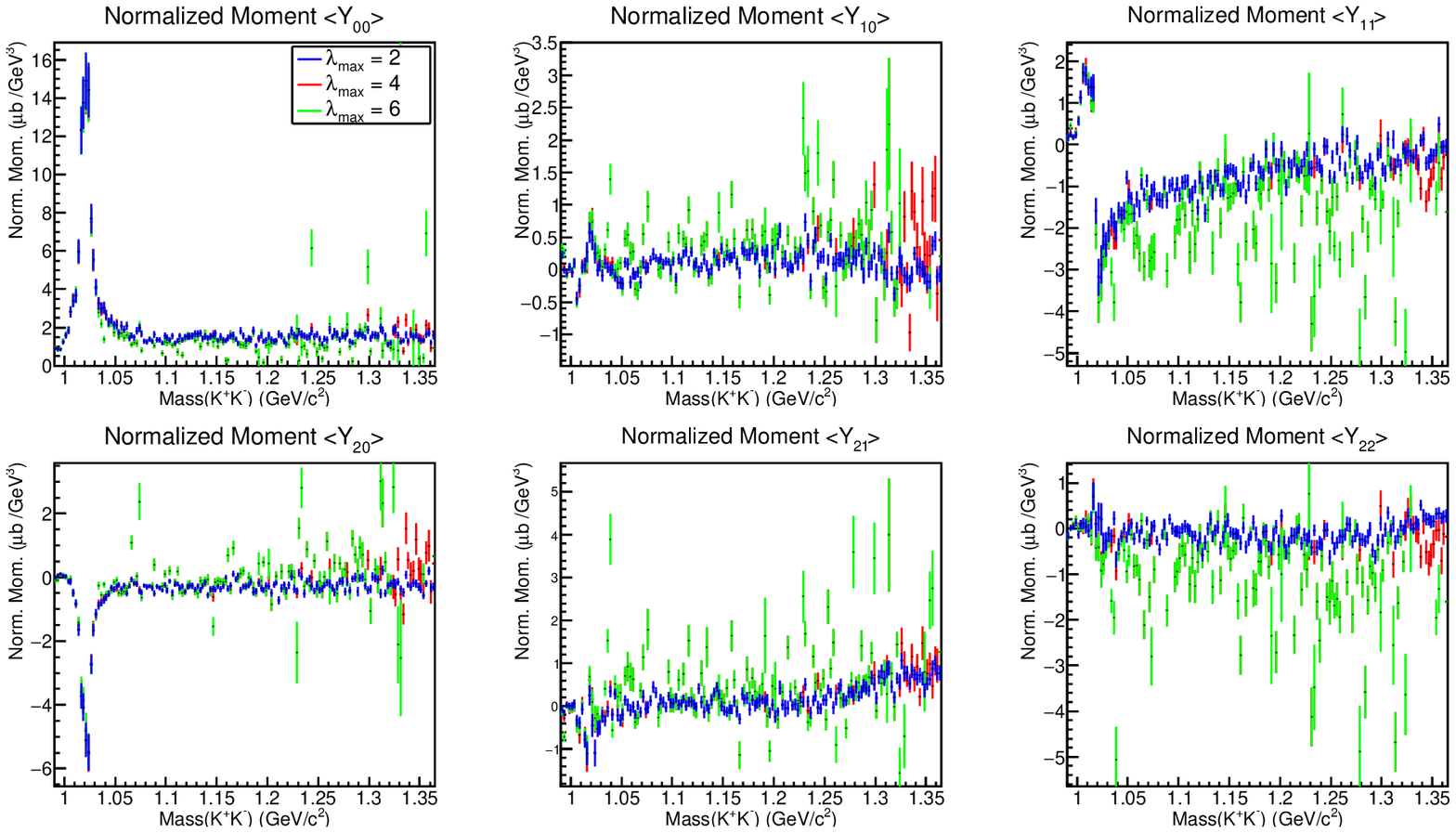}
\caption{Efficiency-corrected, normalized $\langle {Y}_{\lambda \mu}\rangle$ moments from method M1 varying $\lambda_{max}$. $\langle {Y}_{00}\rangle$ corresponds to the normalized cross section.}
\label{fig:cutoff3}
\end{figure*}

Fig. \ref{fig:cutoff3} shows results from method M1 in which the intensity was parametrized by moments and the likelihood was maximized in one energy and $t$ bin ($3.0\le E \le 3.8 \mbox{ GeV}$, $0.6 \le -t \le 0.7 \mbox{ GeV}^2$). $\lambda_{max}$ was varied from 2 up to $\lambda_{max} = 6$. The fits became unstable as the number of free parameters increases to $\lambda_{max}=6$. 

\begin{figure}
\includegraphics[width=3.5in]{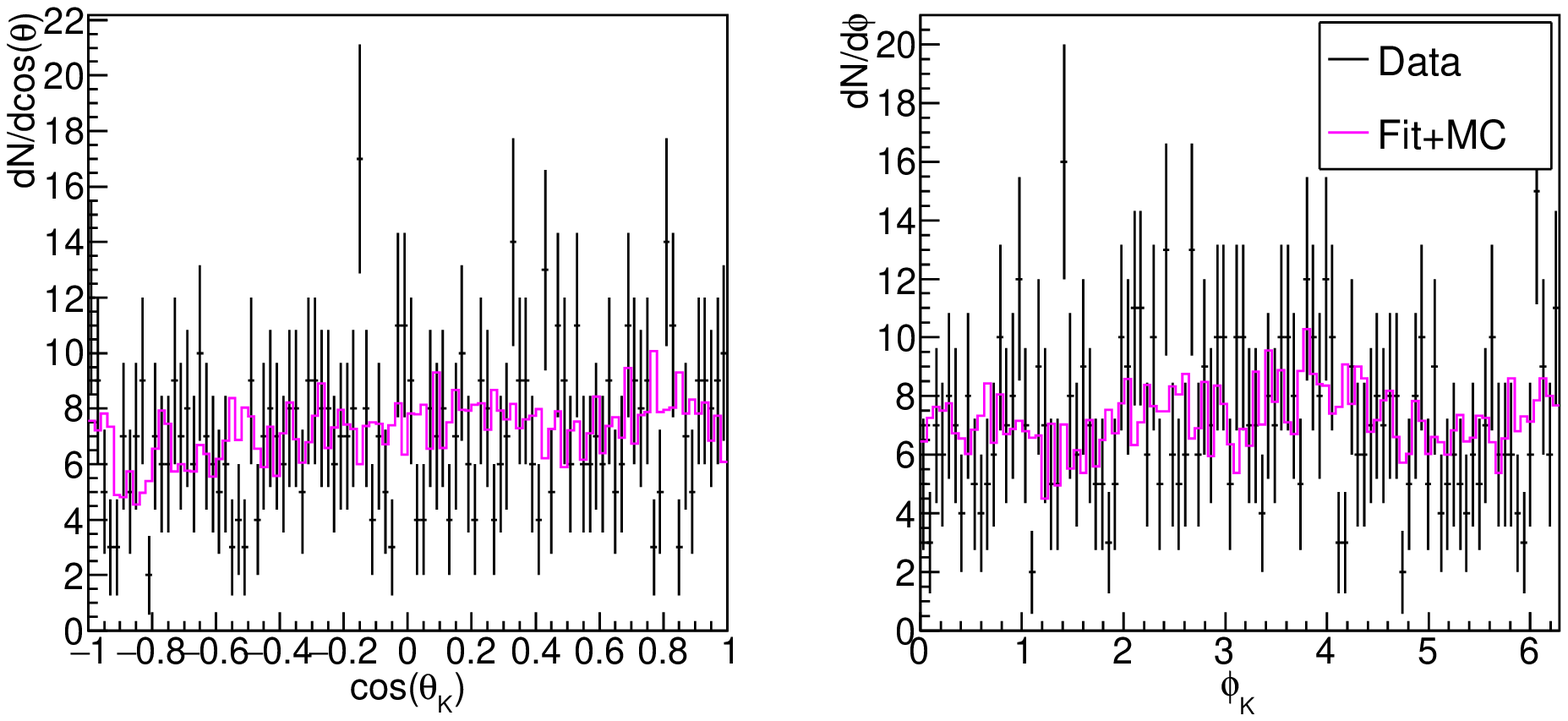}
\includegraphics[width=3.5in]{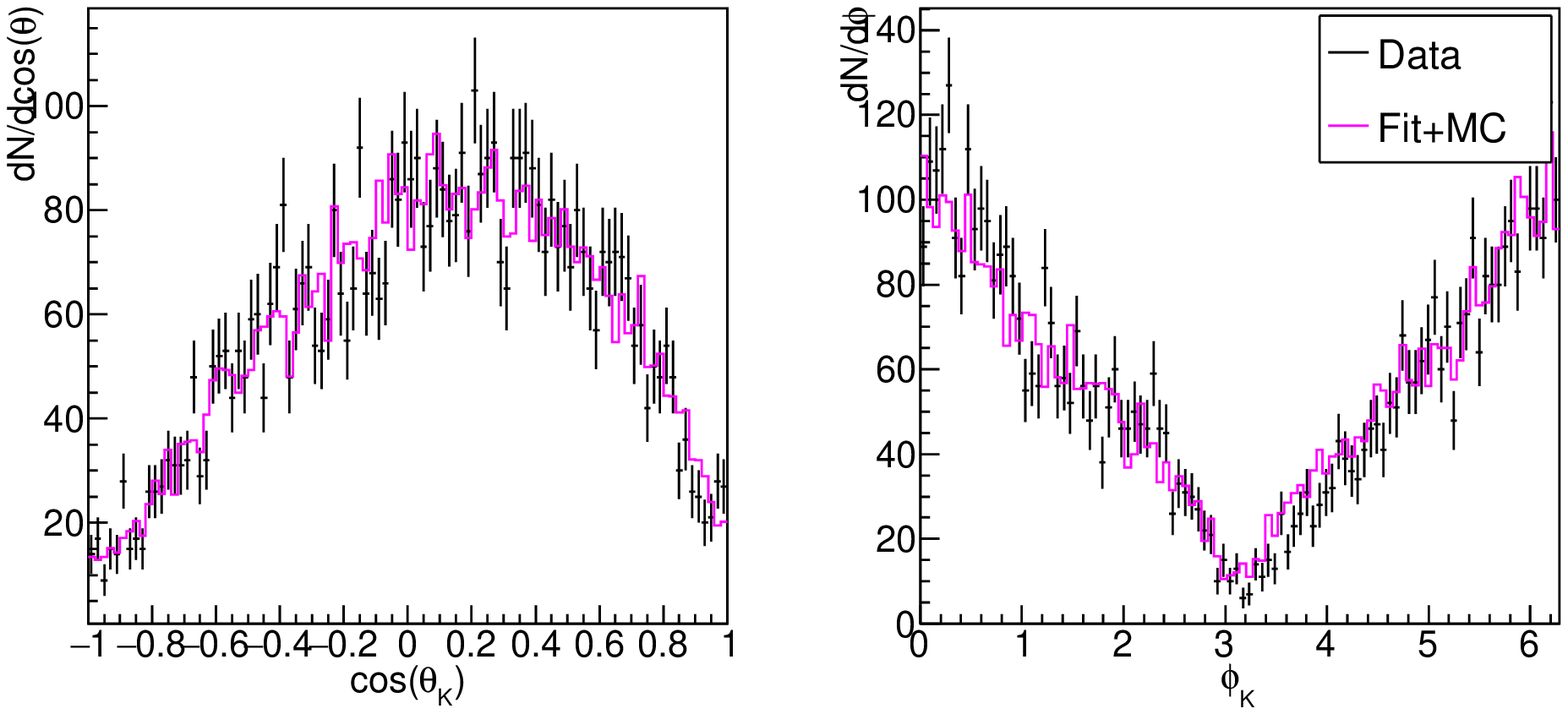}
\includegraphics[width=3.5in]{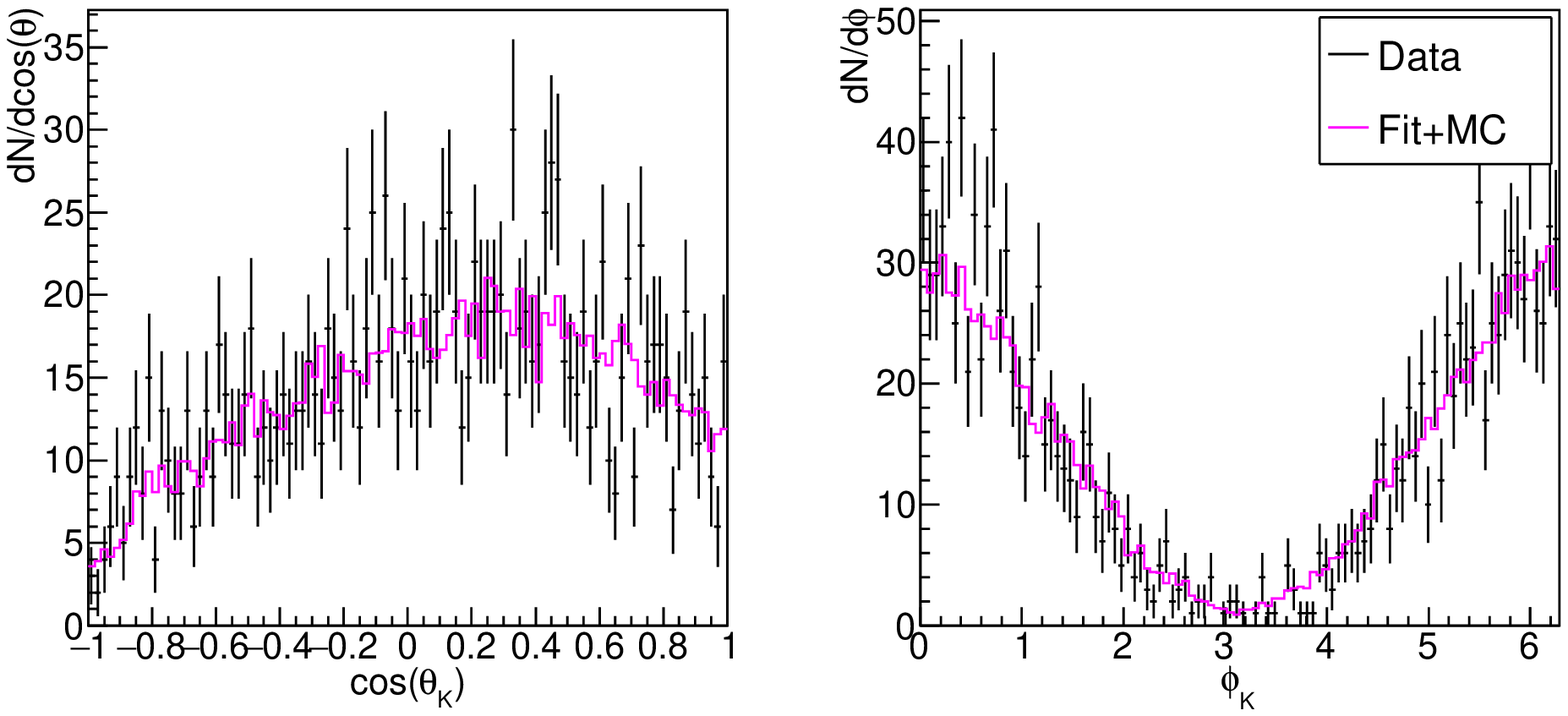}
\caption{Comparison of helicity angles for  $0.4 \le -t \le 1.0$  GeV in the $f_0(980)$ mass region ($M_{K^+K^-} = 0.995 \pm .01$  GeV) (top),  in the $\phi$ mass region ($M_{K^+K^-} = 1.0275 \pm .01$  GeV) (middle),  and above the $\phi$ meson mass region ($M_{K^+K^-} = 1.0575 \pm .01$ GeV) (bottom) for the  measured data (black) and the results reconstructed
      from the fit (purple) using $\lambda_{max}=4$.}
\label{fig:angles}
\end{figure}

\begin{figure}
\includegraphics[width=3.5in]{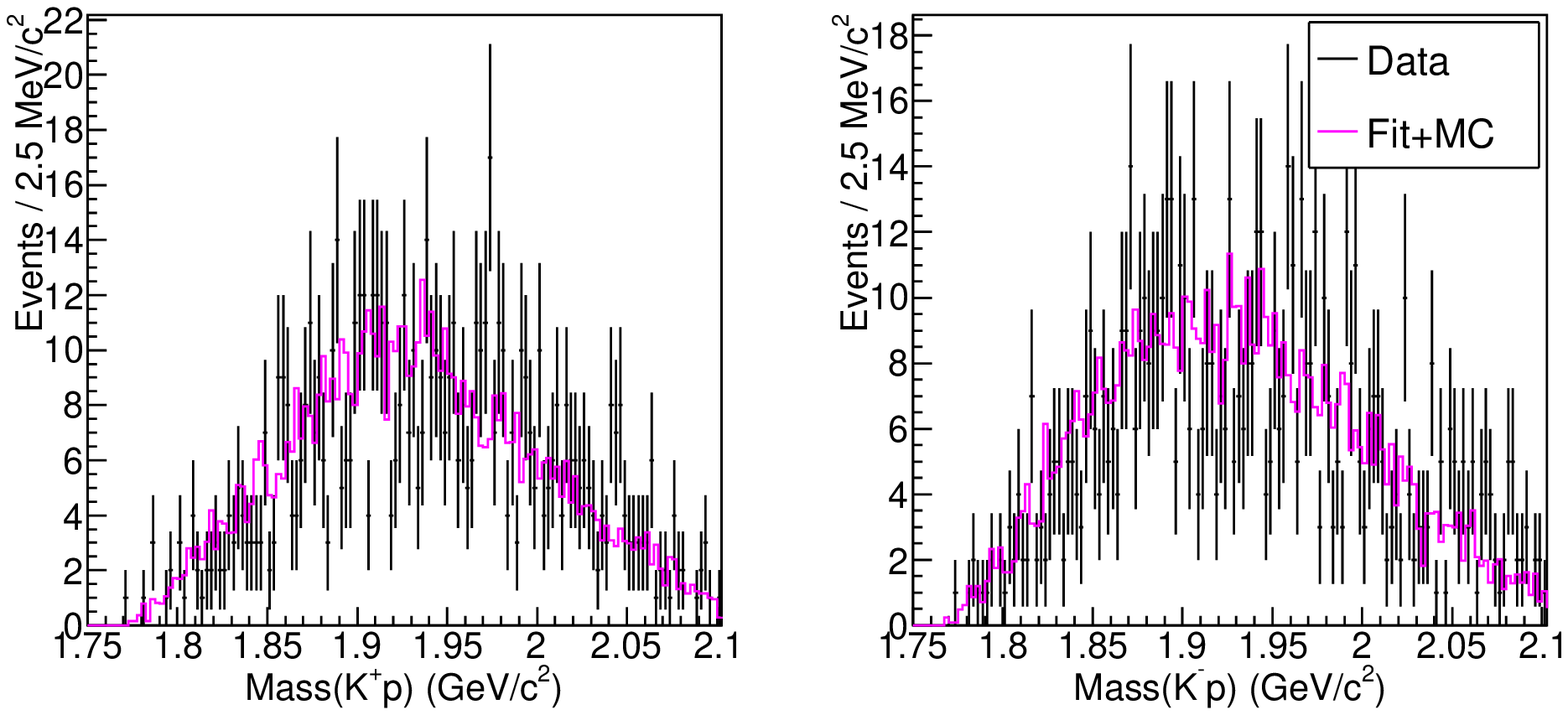}
\includegraphics[width=3.5in]{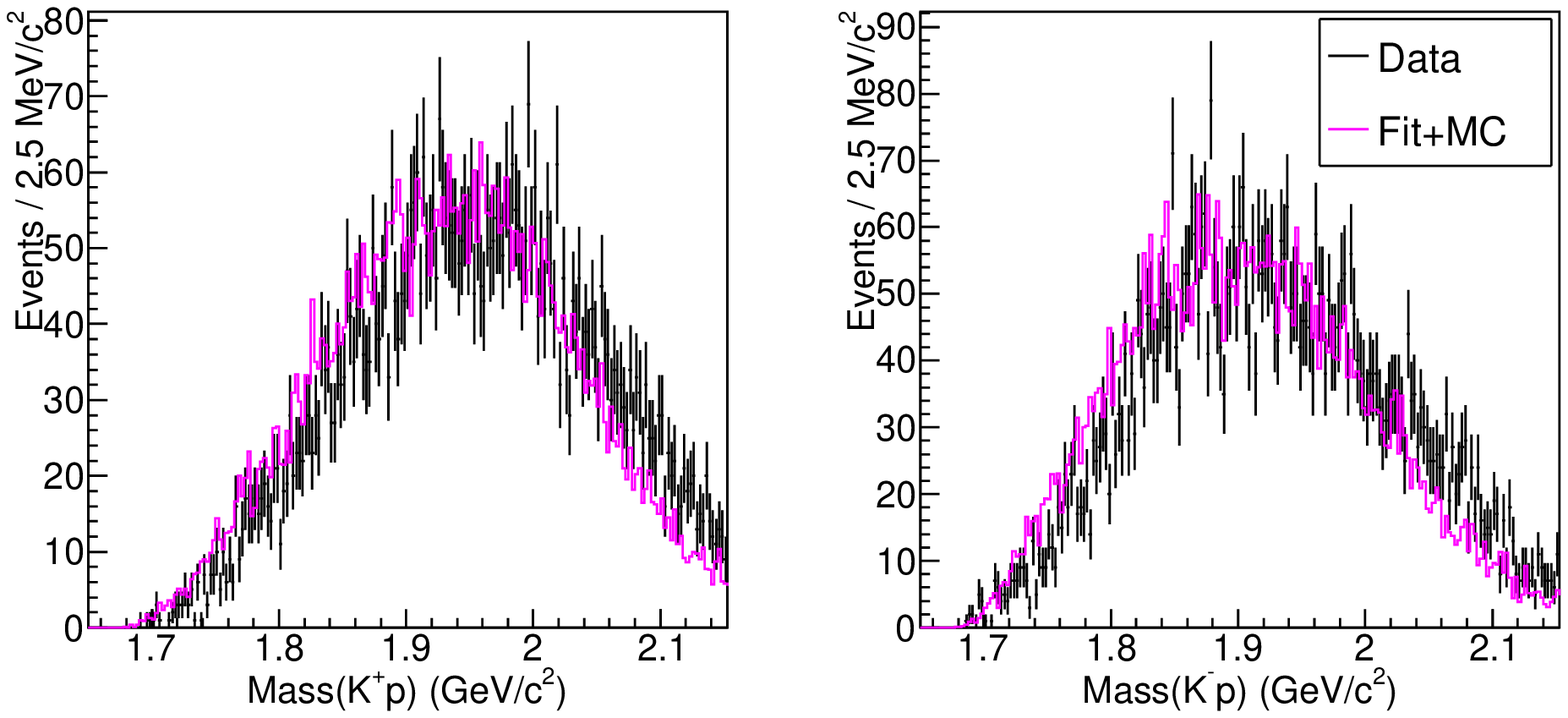}
\includegraphics[width=3.5in]{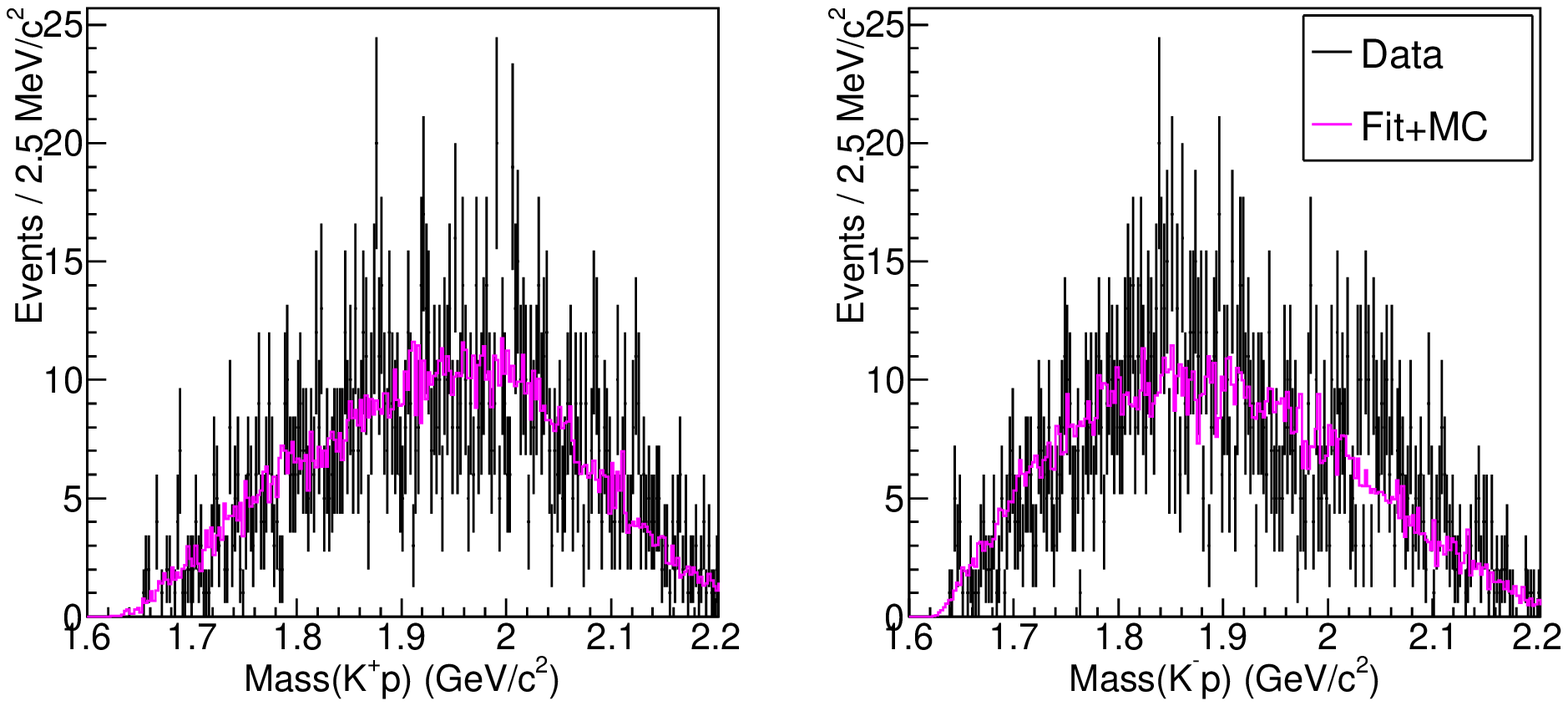}
\caption{Comparison of kaon-nucleon invariant mass distributions for $0.4 \le -t \le 1.0$  GeV in the $f_0(980)$ mass region ($M_{K^+K^-} = 0.995 \pm .01$  GeV) (top),  in the $\phi$ mass region ($M_{K^+K^-} = 1.0275 \pm .01$  GeV) (middle),  and outside of the $\phi$ meson mass region ($M_{K^+K^-} = 1.0575 \pm .01$  GeV) for the measured data (black) and the results reconstructed from the fit procedure (purple) using $\lambda_{max}=4$.}
\label{fig:mass}
\end{figure}

The $\lambda_{max}=4$ fit reproduced the  main features of the data in the region of interest ($M_{K^+K^-} \le 1.1$  GeV). We compare the  helicity angles and invariant masses in Fig. \ref{fig:mass} and Fig.~ \ref{fig:angles} between data and reconstruction from the fit results   (plotting the average of methods M1 and M2) for three different $M_{K^+K^-}$ intervals ($M_{K^+K^-} = 0.995 \pm .01$  GeV $M_{K^+K^-} = 1.0275 \pm .01$    GeV, $M_{K^+K^-} = 1.0575 \pm .01$   GeV). The rationale for this choice of mass regions is as follows. The first region lies to the left of the $\phi$ peak, the second is directly on the peak where the signal is dominated by the $\phi$, and the third region is to the right of the $\phi$ peak.  In the first mass region shown on the top of the figures, a large momentum transfer range ($0.4  \le -t \le 1.0 \mbox{ GeV}^2$) was integrated over to obtain an appreciable number of events. In general, it was found as expected,  that the reconstructed distributions from smaller bin sizes in $t$ and $E$ better reproduce the data. 

The helicity angle distributions reproduced from the fits are in good agreement with the data.
There is a similarity in the $\phi_{K}$ helicity angular distributions between events in the second ($M_{K^+K^-} = 1.0275 \pm .01$  GeV) and third ($M_{K^+K^-} = 1.0575 \pm .01$ GeV) mass range. This is counterintuitive because the angular distribution for $M_{K^+K^-} = 1.0275$  GeV resembles
a $P$-wave signal as expected, but the angular distribution in Fig.
\ref{fig:angles}, which is away from the $\phi$ peak ($M_{K^+K^-} = 1.0575$  GeV), looks similar. We found this can be attributed to the CLAS detector acceptance and not to the presence of a large 
$P$-wave in the third mass interval. The accepted Monte Carlo events, with primary events generated from a flat phase-space distribution, also takes the same form as the data in this region due to the detector acceptance. The shape of the $\phi_{K}$ angular distribution from the data outside of the $\phi$ meson mass region can therefore be explained by the angular dependence of the detector acceptance.

The invariant mass distributions of the data are also described well by the fit. The two regions away from the $\phi$ are shown in the top and bottom plots of  Fig.~\ref{fig:mass}. The kaon-nucleon mass distributions directly on the $\phi$ peak (middle plots) are consistent within one sigma, except for just a few bins.

\section{Parametrization of individual $K^+K^-$ amplitudes }\label{appendix:amp}
\noindent 
We restricted our analysis to waves with $M \le 1$ and partial waves up to $L=1$ waves. 

\subsubsection{$P-$wave}

 The $P$-waves were constructed based on the 
  model of elastic $K^+K^-$ photoproduction developed in 
    \cite{Lesniak:2003}. The model assumes that the $\phi(1020)$ resonance is produced by a soft Pomeron exchange, which leads to an almost purely imaginary amplitude at small momentum transfers. The $K^+ K^-$ effective mass distribution is described 
     by the relativistic Breit-Wigner formula
     \begin{equation}
BW(M_{K^+K^-}) = \frac{1}{M_{\phi}^2 - M_{K^+K^-}^2 - iM_{\phi} \Gamma_{\phi}},
\label{BW}
\end{equation}
 with $M_\phi$ and $\Gamma_{\phi}$ being the $\phi$ meson mass and width. 
 Expanding the $P$-wave amplitudes into partial waves,
\begin{equation} 
 f^{1}_{\sigma, \lambda, \lambda '}(s,t,W,\Omega)  = \sum_M  f^{1M}_{\sigma, \lambda, \lambda ' }(s,t,W) Y_{1M}(\Omega),
 \end{equation}
and taking the high energy limit,  $s \gg t$ and $ s \gg M_{K^+K^-}^2$,  the amplitudes derived in \cite{Lesniak:2003}   result in the following helicity partial waves, 
    \begin{equation}
f^{1,1}_{+++}  = f^{1,1}_{+--} \propto s  \sqrt{M_{K^+K^-}^2 - 4m_K^2} 
   BW(M_{K^+K^-}),
\label{param_first}
\end{equation}

\begin{equation}
f^{1,0}_{+++}  = f^{1,0}_{+--} \propto   s \sqrt{-t}  \sqrt{M_{K^+K^-}^2 - 4m_K^2}   BW(M_{K^+K^-}).
\end{equation}
Before comparing with data  we multiplied each of these amplitudes
by a slowly varying function of $M_{K^+K^-}$, 
\begin{equation} 
f(M_{K^+K^-}) = a + b w(M_{K^+K^-}) + c w^2(M_{K^+K^-}) \label{f} 
\end{equation} 
 with $w(z)$ conformally mapping the complex $M_{K^+K^-}^2$ plane cut at $M^2_{KK} = 0$ and $M^2_{KK} = 4m^2_K$ onto a unit circle.  coefficients $a$, $b$, and $c$ are allowed to vary independently for  each helicity amplitude.

\subsubsection{$S-$wave}

The $S-$wave component of the $K^+K^-$ amplitude is parametrized by the double $t-$channel exchange of the $\rho$ and $\omega$ vector mesons as described in \cite{Lesniak:2005}. In the upper meson vertex, a simple meson exchange is used, allowing for an interaction of two produced mesons in the final state. The normal propagator $(t -m_e^2)^{-1}$, where $m_e$ is the mass of the exchanged vector meson, was used at the nucleon vertex. Both the $\pi^+\pi^-$ and $K^+ K^-$ channels were included in the final state interactions. The $S-$wave in the mass region considered is dominated by the $f_0(980)$ and $a_0(980)$ resonances. Each partial wave helicity $S$-wave amplitude was multiplied by the function $f(M_{K^+K^-})$ given  in Eq.~\eqref{f}, which   contains three independent fit 
parameters.


\newpage

\begin{thebibliography}{99}



\bibitem{elsa} C. Wu \textit{et al.}, Eur. Phys. J. \textbf{A 23}, 317 (2005).

\bibitem{Ostrick:2016eig} 
  M.~Ostrick (MAMI Collaboration),
  JPS Conf.\ Proc.\  {\bf 10}, 010004 (2016).
  doi:10.7566/JPSCP.10.010004
\bibitem{Ireland:2017ksn} 
  D.~Ireland (CLAS Collaboration),
  PoS INPC {\bf 2016}, 265 (2017).

\bibitem{Patsyuk:2017imk} 
  M.~Patsyuk (GlueX Collaboration),
  EPJ Web Conf.\  {\bf 138}, 01029 (2017).
  doi:10.1051/epjconf/201713801029


\bibitem{mesonx} M. Battaglieri {\it et al.} JLab approved experiment E12-11-005: {\it Meson Spectroscopy with low $Q^2$ electron scattering in CLAS12}   (2011)  and   A.~Celentano,  Acta Phys.\ Polon.\ Supp.\  {\bf 6} (2013) no.3,  769.

\bibitem{Ballam73} J. Ballam \textit{et al.}, Phys. Rev \textbf{D 7}, 3150 (1973).

\bibitem{Aston80} D.~Aston {\it et al.}, Nucl. Phys. B {\bf 172}, 1 (1980).
\bibitem{Fries78}  D.C. Fries {\it et al.}, Nucl. Phys. B 143, 408 (1978).


\bibitem{Battaglieri:2012zz} 
  M.~Battaglieri,
  Prog.\ Part.\ Nucl.\ Phys.\  {\bf 67}, 603 (2012).
  
  
\bibitem{Pelaez:2015qba} 
  J.~R.~Pelaez,
  Phys.\ Rept.\  {\bf 658}, 1 (2016).  

\bibitem{Kaminski:2006yv}   R.~Kaminski, J.~R.~Pelaez and F.~J.~Yndurain,  Phys.\ Rev.\  D {\bf 74}, 014001 (2006)
\bibitem{Caprini:2005zr}   I.~Caprini, G.~Colangelo and H.~Leutwyler,  Phys.\ Rev.\ Lett.\  {\bf 96}, 132001 (2006).
\bibitem{Kaminski:2006qe}  R.~Kaminski, J.~R.~Pelaez and F.~J.~Yndurain,  Phys.\ Rev.\  D {\bf 77}, 054015 (2008).


  \bibitem{Dai:2014zta} 
  L.~Y.~Dai and M.~R.~Pennington,
  Phys.\ Rev.\ D {\bf 90}, no. 3, 036004 (2014).
  
 \bibitem{Briceno:2017qmb} 
  R.~A.~Briceno, J.~J.~Dudek, R.~G.~Edwards and D.~J.~Wilson,
  Phys.\ Rev.\ D {\bf 97}, no. 5, 054513 (2018).
  
 
\bibitem{Behrend} H. -J. Behrend \textit{et al.}, Nucl. Phys. \textbf{B 144}, 22 (1978).

\bibitem{Barber} D. P. Barber \textit{et al.}, Z. Phys \textbf{C 12}, 1 (1982).
\bibitem{f0-clas} M.Battaglieri \textit{et al.} (CLAS Collaboration), Phys. Rev. Lett.  \textbf{102}, 102001 (2009).
\bibitem{2pi-clas} M.~Battaglieri {\it et al.}  (CLAS Collaboration), Phys.\ Rev.\  D {\bf 80}, 072005 (2009).
\bibitem{CLAS} B.A. Mecking {\it et al.}, Nucl. Instr. and Meth. A{\bf 503}, 513  (2003).
\bibitem{SO99}  D. I. Sober {\it et al.}, Nucl. Instr. and Meth. {\bf A440}, 263  (2000).
\bibitem{tag-abs_cal} S. Stepanyan {\it et al.}, Nucl. Instr. and Meth. A{\bf 572}, 654  (2007).
\bibitem{devita} R. De Vita {\it et al.}  (CLAS Collaboration), Phys.\ Rev.\  D {\bf 74}, 032001 (2006).
\bibitem{DC}    M.D. Mestayer    {\it et al.}, Nucl. Instr. and Meth. {\bf A449}, 81 (2000).
\bibitem{Sm99}  E.S. Smith {\it et al.}, Nucl. Instr. and Meth. {\bf A432}, 265  (1999).
\bibitem{ST} Y.G. Sharabian {\it et al.}, Nucl. Instr. and Meth. {\bf A556}, 246  (2006).


\bibitem{Chung:1997qd} 
  S.~U.~Chung,
  Phys.\ Rev.\ D {\bf 56}, 7299 (1997).



\bibitem{sal-note} S. Lombardo  CLAS-Analysis Note 2017 - 007 
 {\em https://misportal.jlab.org/ul/Physics/Hall-B/clas/viewFile.cfm/2017-007.pdf?documentId=773}
\bibitem{jlab-db} JLab Experiment CLAS Database \texttt{http://clasweb.jlab.org/physicsdb/intro.html}
\bibitem{dhuram-db} The Durham HEP Databases \texttt{http://durpdg.dur.ac.uk/}
\bibitem{phi-clas-2014} B.~Dey  {\it et al.}  (CLAS Collaboration),  Phys.\ Rev.\  C {\bf 89}, 055208 (2014).
\bibitem{Irving:1977ea} 
  A.~C.~Irving and R.~P.~Worden,
  Phys.\ Rept.\  {\bf 34}, 117 (1977).

\bibitem{Lesniak:2003} 
  L.~Lesniak and A.~P.~Szczepaniak,
  Acta Phys.\ Polon.\ B {\bf 34}, 3389 (2003)
  [hep-ph/0304007].

\bibitem{Lesniak:2005} 
  L.~Bibrzycki, L.~Lesniak and A.~P.~Szczepaniak,
  Eur.\ Phys.\ J.\ C {\bf 34}, 335 (2004)
  doi:10.1140/epjc/s2004-01724-6
  [hep-ph/0308267].






\bibitem{Bibrzycki:2013pja} 
  L.~Bibrzycki and R.~Kaminski,
  Phys.\ Rev.\ D {\bf 87}, no. 11, 114010 (2013).

\end{thebibliography}
\end{document}